\documentclass[aps,prd,preprint,floats,epsf,superscriptaddress,nofootinbib]{revtex4}

\usepackage{graphicx} 
\usepackage{amsmath,amssymb,amsfonts,mathtools} 
\usepackage{mathrsfs}
\usepackage{slashed} 
\usepackage{braket} 
\usepackage{indentfirst}
\usepackage{xcolor}
\usepackage{dsfont}
\usepackage{ulem}
\usepackage{dcolumn}
\usepackage{bm}
\usepackage{bbm}

\usepackage[hyperfootnotes=true]{hyperref}
\allowdisplaybreaks

\begin{document}

\title{{A  Simple Model of Dark Matter and CP Violation}}

\vspace*{0.2cm}
\author{\vspace{0.5cm}Ting-Kuo Chen}
\affiliation{Department of Physics and Center for Theoretical Physics, National Taiwan University, Taipei 10617, Taiwan}
\affiliation{Physics Division, National Center for Theoretical Sciences, Taipei 10617, Taiwan}

\author{Cheng-Wei Chiang}
\affiliation{Department of Physics and Center for Theoretical Physics, National Taiwan University, Taipei 10617, Taiwan}
\affiliation{Physics Division, National Center for Theoretical Sciences, Taipei 10617, Taiwan}

\author{Ian Low\vspace{0.5cm}}
\affiliation{High Energy Physics Division, Argonne National Laboratory, Lemont, IL 60439, USA}
\affiliation{\mbox{Department of Physics and Astronomy, Northwestern University, Evanston, IL 60208, USA}\vspace{0.5cm}}

\begin{abstract}
\vspace{0.5cm}
We propose a simple model of dark matter and CP violation and consider the associated triple and quadruple productions of 125 GeV Higgs bosons at the Large Hadron Collider (LHC). In the model, the dark matter is a vector-like dark fermion $(\bar{\chi}, \chi)$ interacting with the Standard Model only through a  complex messenger scalar $S$ which is an electroweak singlet. New sources of CP violation reside in the most general scalar potential involving the  doublet $H$ and the  singlet $S$, as well as in the dark Yukawa coupling between $S$ and $(\bar{\chi}, \chi)$. We study current experimental constraints from Higgs measurements, searches for new scalars at the LHC, precision electroweak measurements, EDM measurements, dark matter relic density, as well as direct and indirect detections of dark matter. A smoking-gun signature of CP violation could come from the Higgs-to-Higgs decays, $h_3\to h_2h_1$, where $h_3/h_2/h_1$ are the heaviest scalar, second heaviest scalar and the SM-like 125-GeV Higgs, respectively. Taking into account other Higgs-to-Higgs decays, such as $h_3\to 2h_2$ and $h_3/h_2\to 2h_1$, then gives rise to novel $3h_1$ and $4h_1$ final states, which have yet to be searched for experimentally. We present four benchmarks and show the event rates for $3h_1$ and $4h_1$ final states could be as large as ${\cal O}(10)\ {\rm fb}$ and ${\cal O}(1)\ {\rm fb}$, respectively, at the 14-TeV LHC. This work opens up a new frontier of searching for triple and quadruple Higgs bosons at a high energy collider.
\end{abstract}

\maketitle

\newpage

\section{Introduction}\label{sec:1}

Dark matter and CP violation (CPV) are two of the most pressing puzzles in physics nowadays. Both  relate to our own being in the Universe: dark matter is necessary for structure formation and CPV is a required condition for the observed matter-antimatter asymmetry. In particular, there is no cold dark matter candidate in the Standard Model (SM) of particle physics and the amount of CPV in the SM is insufficient to generate the observed baryon asymmetry. Consequently, both problems hint at the presence of new physics beyond the SM.

In this work we propose a simple extension of the SM to accommodate the  dark matter and new sources of CPV, by including a vector-like dark fermion $(\bar{\chi},\chi)$ as the dark matter and a complex singlet scalar $S$ as the messenger mediating interactions between the SM and the dark matter. The most general scalar potential involving the Higgs doublet $H$ and the singlet $S$ contains several new sources of CPV, as does the dark Yukawa coupling between $S$ and $(\bar{\chi},\chi)$. The complex singlet scalar extended SM has been studied in many contexts, such as the CPV, electroweak baryogenesis (EWBG), electroweak phase transition (EWPT), and scalar dark matter~\cite{Barger:2008jx,Chiang:2008ud,AlexanderNunneley:2010nw,Barger:2010yn,Gonderinger:2012rd,Gabrielli:2013hma,Jiang:2015cwa,Darvishi:2016gvm,Darvishi:2016fwo,Chiang:2017nmu,Chiang:2019oms,Robens:2019kga}. Our model is distinct in that i) we do not impose any discrete symmetries in the scalar potential, ii) the dark matter candidate is a vector-like dark fermion, instead of a component of the singlet scalar, and iii)  new sources of CPV are confined in the scalar potential and the dark Yukawa coupling.\footnote{If the complex singlet scalar $S$ only couples to the SM Higgs doublet $H$, its CP-property is not well-defined since the  transformations $S\xrightarrow{\rm CP}S$ and $S\xrightarrow{\rm CP}S^*$ are both allowed~\cite{Robens:2019kga,Ivanov:2017dad}; introducing the dark fermion allows us to define the CP-property of $S$ through the dark Yukawa coupling.}  We also do not introduce higher dimensional operators beyond the renormalizable level.

An important aspect of our model is the consideration of the ``alignment limit'' ~\cite{Carena:2013ooa,Carena:2014nza,Carena:2015moc}, where properties of the 125-GeV Higgs boson have been measured to be closely aligned with those of a SM Higgs boson, and its interplay with the CPV in the scalar sector. Previously this interplay was studied in the context of complex two-Higgs doublet models (C2HDM)~\cite{Grzadkowski:2014ada,Grzadkowski:2018ohf,Kanemura:2020ibp,Low:2020iua}. In particular, Ref.~\cite{Low:2020iua} pointed out the Higgs-to-Higgs decay in $h_3\to h_2h_1$ and the resulting triple Higgs final state as a novel signature for CPV in the C2HDM and presented benchmarks where the triple Higgs final states could be discovered at the High-Luminosity Large Hadron Collider (HL-LHC). However, the C2HDM model is severely constrained by the electric dipole moment (EDM) measurements and the triple scalar coupling mediating the $h_3\to h_2h_1$ decay is suppressed near the exact alignment limit \cite{Low:2020iua}. We will see that in our complex singlet scalar extended model, there is no new physics contribution to the EDM and the particular scalar coupling in $h_3\to h_2h_1$ is not suppressed near the alignment limit. Furthermore, including the other Higgs-to-Higgs decays in $h_3\to2h_2$ and $h_2/h_3\to 2h_1$, there is not only the triple Higgs but also the quadruple Higgs final states!

After performing a comprehensive study on current experimental constraints from Higgs measurements, searches for new scalars at the LHC, precision electroweak measurements, electron EDM measurements, dark matter relic density, as well as direct and indirect detections of dark matter, we present four benchmarks and consider the collider phenomenology. Two of the benchmarks are chosen to allow for a significant $3h_1$ production, while the other two have both $3h_1$ and $4h_1$ productions. Moreover, in two benchmarks the dark matter relic density agrees with current measurements.

This paper is organized as follows. In Sec.~\ref{sec:2}, we introduce the complex singlet scalar extended model with the dark matter (CPVDM model), and identify the CP-conserving (CPC) and the general CPV scenarios.  In Sec.~\ref{sec:3}, we study the experimental constraints from the LHC Higgs measurements, electroweak oblique corrections, direct searches for heavy scalars, the DM relic density and its direct {and indirect} search bounds.  In Sec.~\ref{sec:4}, we present the four benchmarks and study the corresponding $3h_1/4h_1$ decays at the LHC, assuming a centre-of-mass  energy at $14$~TeV.  Finally in Sec.~\ref{sec:5}, we conclude our study and propose future prospects. We also provide two appendices: Appendix \ref{sec:a} contains the full list of scalar couplings and Appendix \ref{sec:b} presents the formulas needed for computing the electroweak oblique corrections.

\section{The Model}\label{sec:2}

In addition to the SM Higgs doublet denoted by $H$ with hypercharge $1/2$\footnote{We adopt the hypercharge convention $Q = Y+T_3$.}, we introduce a complex scalar singlet $S$ with hypercharge $0$. The most general renormalizable scalar potential $V(H,S)$ consistent with required symmetries is given by~\cite{Barger:2008jx}\footnote{We modify the parameter convention of the SM Higgs potential in Ref.~\cite{Barger:2008jx} by setting $m^2/2\to\mu^2$ and $\lambda/4\to\lambda$.}
\begin{equation}
\begin{aligned}
	V(H,S) &= \mu^2H^\dagger H + \lambda(H^\dagger H)^2+\frac{\delta_1}{4}H^\dagger HS+\frac{\delta_2}{2}H^\dagger H\vert S\vert^2 + \frac{\delta_3}{4}H^\dagger HS^2 
	\\
	&\quad +\frac{b_1}{4}S^2 +\frac{b_2}{2}\vert S\vert^2+\frac{c_1}{6}S^3+\frac{c_2}{6}S\vert S\vert^2 
	+\frac{d_1}{8}S^4+\frac{d_2}{4}\vert S\vert^4+\frac{d_3}{8}S^2\vert S\vert^2 + \mbox{H.c.}
	~,
\end{aligned}
\label{L:Scalar}
\end{equation}
where the couplings $\delta_1,\delta_3,b_1,c_1,c_2,d_1,d_3$ are generally complex, and the term linear in the $S$ field has been removed without loss of generality.  While in most complex singlet scalar extended Standard Model studies, extra symmetries are often imposed to simplify the potential or to have a DM candidate~\cite{Bento:1991ez,Branco:2003rt,Barger:2008jx,Costa:2014qga,Jiang:2015cwa,Darvishi:2016gvm,Darvishi:2016tni,Darvishi:2016fwo}, we keep the potential as general as possible without imposing further symmetries in this work. The DM will arise out of the vector-like fermion (VLF) $\chi$, which we will discuss later.

With the SM Higgs vacuum expectation value (VEV) $v\approx246$~GeV and assuming that $S$ attains a VEV, $\langle S\rangle=v_s\exp(i\xi)$, where $\xi$ is a generally nonzero phase, we parametrize the two scalars as
\begin{equation}
	H=\begin{pmatrix}
		G^+ \\
		\frac{1}{\sqrt{2}}(v+\phi_1+iG^0)
	\end{pmatrix},~S=\frac{1}{\sqrt{2}}(v_s+\phi_2+ia)e^{i\xi} ~, \label{fields}
\end{equation}
where $G^+$ and $G^0$ are the Goldstone bosons to be ``eaten'' by the weak gauge bosons.  With the freedom to rephase $S$, we choose to make $\langle S\rangle$ real and absorb $\xi$ into the the Lagrangian parameters, resulting in the redefinitions:
\begin{align}
\begin{split}
&
	\theta_{\delta_1}+\xi\to\theta_{\delta_1}
	~,~
	\theta_{c_2}+\xi\to\theta_{c_2}
	~,~
	\theta_{\delta_3}+2\xi\to\theta_{\delta_3}
	~,~
	\theta_{b_1}+2\xi\to\theta_{b_1}~,
	\\
&
	\theta_{d_3}+2\xi\to\theta_{d_3} 
	~,~
	\theta_{c_1}+3\xi\to\theta_{c_1} 
	~,~
	\theta_{d_1}+4\xi\to\theta_{d_1} 
	~,
\end{split}
\label{eq:rephase}
\end{align}
where we have parametrized the complex parameters in the scalar potential as $x=\vert x\vert e^{i\theta_{x}}$. From this reasoning it is clear that the conditions for CP invariance in the scalar sector is such that all phases in Eq.~(\ref{eq:rephase}) now vanish upon a phase rotation in $S$ to make $\langle S\rangle $ real:
\begin{equation}
\label{eq:cpcondi}
{\rm \bf CP\ Invariance} :\  \xi = -\theta_{\delta_1}= -\theta_{c_2} = -\frac12 \theta_{\delta_3} = -\frac12 \theta_{b_1} = -\frac12 \theta_{d_3} = -\frac13 \theta_{c_1}  = -\frac14 \theta_{d_1} \ . 
\end{equation}

Next, we introduce singlet VLF fields $\chi_{L,R}$, which only couples to the singlet scalar $S$. In this sense $S$ is a messenger field between the dark sector $\chi$, which has odd parity under a $\mathbb{Z}_2$ symmetry, and the SM.  The Yukawa interactions involving $\chi$ and $S$ are given by
\begin{align}
	\mathcal{L}_{NP} = -\lambda_\chi  S\overline{\chi}_L\chi_R + \mbox{H.c.} ~,
\end{align}
where  $\lambda_\chi$ can be made real by a chiral phase rotation on $\chi$.  In the end, 
\begin{align}
	\mathcal{L}_{NP} = -\frac{\lambda_\chi}{\sqrt{2}}\overline{\chi}(v_s+\phi_2+i\gamma_5a)\chi ~,
\end{align}
where $\phi_2$ and $a$ are CP-even and CP-odd, respectively. For simplicity we assume the VLF  receives all of its mass from the singlet VEV:\footnote{In general, we could include a Dirac mass term for the dark matter, which would not change the phenomenology other than giving rise to an extra free parameter.}
\begin{equation}
m_\chi=\lambda_\chi v_s/\sqrt{2}\ .
\label{eq:dmmass}
\end{equation} 
 We impose a $\mathbb{Z}_2$ symmetry under which only $\chi_{L,R}$ have odd parity, making it a DM candidate.   The field contents of our CPVDM model and the corresponding quantum numbers are summarized in Table~\ref{table:charge}.

\begin{table}[t]
\centering
\begin{tabular}{>{\centering\arraybackslash}p{1.5cm}||>{\centering\arraybackslash}p{1.5cm}|>{\centering\arraybackslash}p{1.5cm}|>{\centering\arraybackslash}p{1.5cm}|>{\centering\arraybackslash}p{1.5cm}}
\toprule
Field & $SU(3)_C$ & $SU(2)_L$ & $U(1)_Y$ & $\mathbb{Z}_2$ \\
\toprule
$Q_L$ & $3$ & $2$ & $\frac{1}{6}$ & $+$ \\
\colrule
$u_R$ & $3$ & $1$ & $\frac{2}{3}$ & $+$ \\
\colrule
$d_R$ & $3$ & $1$ & $-\frac{1}{3}$ & $+$ \\
\colrule
$L_L$ & $1$ & $2$ & $-\frac{1}{2}$ & $+$ \\
\colrule
$\ell_R$ & $1$ & $1$ & $-1$ & $+$ \\
\colrule
$\chi_{L,R}$ & $1$ & $1$ & $0$ & $-$ \\
\toprule
$H$ & $1$ & $2$ & $\frac{1}{2}$ & $+$ \\
\colrule
$S$ & $1$ & $1$ & $0$ & $+$ \\
\botrule
\end{tabular}
\caption{\label{table:charge} Field contents of the CPVDM model and their corresponding charge assignments.}
\end{table}


Using the redefined parameters in Eq.~(\ref{eq:rephase}), the minimization of the scalar potential in Eq.~(\ref{L:Scalar}) gives the following conditions:
\begin{align}
\begin{split}
&
{\frac{\partial V}{\partial \phi_1}\Big\vert_{0}}=4\mu^2+4\lambda v^2+\sqrt{2}v_s{\rm Re}\delta_1+v_s^2\big(\delta_2+{\rm Re}\delta_3\big) = 0 
~, \\
&
{\frac{\partial V}{\partial \phi_2}\Big\vert_{0}}=\sqrt{2}v^2{\rm Re}\delta_1+2v^2v_s\big(\delta_2 +{\rm Re}\delta_3\big)+4v_s\big({\rm Re} b_1+b_2\big)+2\sqrt{2}v_s^2{\rm Re}\big( c_1+ c_2\big)
\\
& \qquad \quad
+2v_s^3\big[d_2+{\rm Re}(d_1+ d_3)\big] = 0  
~, \\
&
{\frac{\partial V}{\partial a}\Big\vert_{0}}=3\sqrt{2}v^2{\rm Im}\delta_1+6v^2v_s{\rm Im}\delta_3+12v_s{\rm Im} b_1+2\sqrt{2}v_s^2{\rm Im}\big(3c_1+c_2\big)
\\
& \qquad \quad
+3v_s^3{\rm Im}\big(2d_1+d_3\big) = 0  
~.
\end{split}
\label{minimize}
\end{align}
The entry of the $3\times 3$ mass-squared matrix $M^2$ in the $(\phi_1,\phi_2,a)^T$ basis is given by
\begin{equation}
\label{eq:mdetail}
\begin{aligned}
	M^2_{11} &= \mu^2+3\lambda v^2+\frac{\sqrt{2}}{4}v_s{\rm Re}\delta_1+\frac{1}{4}v_s^2\big(\delta_2+{\rm Re}\delta_3\big) ~, \\
	M^2_{22} &= \frac{1}{2}\big({\rm Re} b_1+b_2\big)+\frac{1}{4}v^2\big(\delta_2+{\rm Re}\delta_3\big)+\frac{\sqrt{2}}{2}v_s{\rm Re}\big(c_1+c_2\big)+\frac{3}{4}v_s^2\big[d_2+{\rm Re}\big(d_1+d_3\big)\big] ~, \\
	M^2_{33} &= -\frac{1}{2}\big({\rm Re} b_1-b_2\big)+\frac{1}{4}v^2\big(\delta_2-{\rm Re}\delta_3\big)-\frac{\sqrt{2}}{6}v_s{\rm Im}\big(3c_1-c_2\big)-\frac{1}{4}v_s^2{\rm Re}\big(3d_1-d_2\big) ~, \\
	M^2_{12} &= \frac{\sqrt{2}}{4}v{\rm Re}\delta_1+\frac{1}{2}vv_s\big(\delta_2+{\rm Re}\delta_3\big) ~, \\
	M^2_{13} &= -\frac{\sqrt{2}}{4}v{\rm Im}\delta_1-\frac{1}{2}vv_s{\rm Im}\delta_3 ~, \\
	M^2_{23} &= -\frac{1}{2}{\rm Im} b_1-\frac{1}{4}v^2{\rm Im}\delta_3-\frac{\sqrt{2}}{6}v_s{\rm Im}\big(3c_1+c_2\big)-\frac{3}{8}v_s^2{\rm Im}\big(2d_1+d_3\big) ~.
\end{aligned}
\end{equation}
The mixing matrix $R$, which relates the physical mass eigenstates to the original basis $(h_3,h_2,h_1)^T\equiv R(\phi_1,\phi_2,a)^T$, involves three Euler angles:
\begin{equation}
\label{eq:mixing}
	R=\begin{pmatrix}
		c_{12} & -s_{12} & 0 \\
		s_{12} & c_{12} & 0 \\
		0 & 0 & 1
	\end{pmatrix}\begin{pmatrix}
		c_{13} & 0 & -s_{13} \\
		0 & 1 & 0 \\
		s_{13} & 0 & c_{13}
	\end{pmatrix}\begin{pmatrix}
		1 & 0 & 0 \\
		0 & c_{23} & -s_{23} \\
		0 & s_{23} & c_{23}
	\end{pmatrix} ~,
\end{equation}
where $s_{ij} \equiv \sin\theta_{ij}$ and $c_{ij} \equiv \cos\theta_{ij}$, with $\theta_{ij}$ being the mixing angles. The ranges of the Euler angles are given according to the Tait-Bryan convention by
\begin{equation}
	\theta_{12}\in[-\pi/2,\pi/2],\quad \theta_{13}\in[-\pi,\pi], \quad \theta_{23}\in[-\pi,\pi] ~.
\end{equation}

What is the alignment condition such that one of the neutral Higgs bosons is exactly SM-like? Since the messenger scalar $S$ is a singlet and does not couple to the electroweak gauge bosons and the SM fermions, the 125-GeV Higgs boson $h_1$ will be SM-like if the 125-GeV mass eigenstate coincides with $\phi_1$, the neutral scalar in $H$. This can be achieved if, in the mass-squared matrix,  $M_{12}^2 = M_{13}^2=0$. From Eq.~(\ref{eq:mdetail}) we see that this leads to the condition:
\begin{equation}
{\rm \bf Alignment\ Condition}:\quad   M_{12}^2 = M_{13}^2=0 \ \ \Leftrightarrow \ \ 
\frac1{\sqrt{2}}\delta_1 +v_s (\delta_2+\delta_3) = 0 \ .
\end{equation}
In terms of the mixing matrix $R$, $h_1$ is aligned with $\phi_1$ if $\theta_{13}=\pi/2$, in which case $h_1$ does not have components in $\phi_2$ or $a$.

In reality we are only able to establish ``approximate'' alignment limit due to the experimental uncertainty. In this regard, we set $\theta_{13}=\frac{\pi}{2}+\epsilon$ with $\epsilon\ll1$. Thus, we have  
\begin{equation}
\begin{aligned}
R	&=\begin{pmatrix}
		-\epsilon c_{12}\ &\  -c_{23}s_{12}-c_{12}s_{23}\ &\ -c_{12}c_{23}+s_{12}s_{23} \\
		-\epsilon s_{12}\ &\  c_{12}c_{23}-s_{12}s_{23}\ &\ -c_{23}s_{12}-c_{12}s_{23} \\
		1 & -\epsilon s_{23} & -\epsilon c_{23}
	\end{pmatrix}  +{\cal O}(\epsilon^2)~.
\end{aligned}\label{R:matrix}
\end{equation}

In the scalar potential in Eq.~(\ref{L:Scalar}), there are 5 real parameters $\{\mu^2, \lambda, \delta_2, b_2, d_2\}$ and 7 complex parameters $\{\delta_1, \delta_3, b_1, c_1, c_2, d_1, d_3\}$. Among the three minimization conditions in Eq.~(\ref{minimize}), two of them can be viewed as the defining relations for $v$ and $v_s$. As such, only one is a constraint among the parameters of the potential. Moreover, we have chosen the phase of $S$ such that its VEV is real. Thus, in the end there are 18 real degrees of freedom in the scalar potential, which we choose to be the following parameters:
\begin{equation}
\label{eq:input}
	\{m_{h_1},m_{h_2},m_{h_3},v,v_s,\epsilon,\theta_{12},\theta_{23},\delta_2,b_2,\vert c_1\vert,\theta_{c_1},\vert c_2\vert,\theta_{c_2},\vert \delta_3\vert,\theta_{\delta_3},\vert d_3\vert,\theta_{d_3}\} ~.
\end{equation}
There is an additional input parameter as the DM mass $m_\chi$ defined in Eq.~(\ref{eq:dmmass}).
Using these input parameters, some trilinear couplings of particular interest can be written as:
\begin{align}
\begin{split}
g_{123} 
=& -\frac{v}{2}{\rm Im}\big[\delta_3e^{-2i(\theta_{12}+\theta_{23})}\big] + \mathcal{O}(\epsilon)
~, \\
g_{223} 
=& -\frac{1}{12v_s}\Big\{ 3s_{12+23}\big[ -8m_{h_2}^2 + 4m_{h_3}^2 + v^2\delta_2 - 4b_2(1+3c_{2(12+23)}) 
\\
& + 3v_s^2s_{2(12+23)}{\rm Im} d_3 \big] - 9v^2{\rm Im}\big(\delta_3e^{-3i(\theta_{12}+\theta_{23})}\big)
\\
& -\sqrt{2}v_s{\rm Im}\big(3c_1e^{-3i(\theta_{12}+\theta_{23})} - 2c_2e^{i(\theta_{12}+\theta_{23})}+3c_2e^{3i(\theta_{12}+\theta_{23})}\big) \Big\} + \mathcal{O}(\epsilon)
	~, \\
g_{112} 
=& \frac{\epsilon}{v}\Big\{\big(2m_{h_1}^2+m_{h_2}^2\big)\sin\theta_{12}-v^2\big[\delta_2\sin\theta_{12}+{\rm Im}\big(\delta_3e^{-i(\theta_{12}+2\theta_{23})}\big)\big]\Big\} + \mathcal{O}(\epsilon^2)
~, \\
g_{113} 
=& \frac{\epsilon}{v}\Big\{\big(2m_{h_1}^2+m_{h_3}^2\big)\cos\theta_{12}-v^2\big[\delta_2\cos\theta_{12}-{\rm Re}\big(\delta_3e^{-i(\theta_{12}+2\theta_{23})}\big)\big]\Big\} + \mathcal{O}(\epsilon^2)
	~,
\end{split}
\label{eq:gijkcoupling}
\end{align}
where we use $g_{ijk(\ell)}$ to denote the trilinear (quartic) coupling among the physical eigenstates  $h_ih_jh_k(h_\ell)$ in the Lagrangian\begin{equation}\begin{aligned}
	V(H,S)&\ni \frac{g_{iii}}{3!}h_i^3 + \frac{g_{iij}}{2!}h_i^2h_j + g_{ijk}h_ih_jh_k + \frac{g_{iiii}}{4!}h_i^4 + \frac{g_{iiij}}{3!}h_i^3h_j + \frac{g_{iijj}}{2!2!}h_i^2h_j^2 + \frac{g_{iijk}}{2!}h_i^2h_jh_k
	~.
\end{aligned}
\end{equation}
All of the scalar couplings are expanded to the first non-vanishing order in $\epsilon$.
A complete list of all the trilinear and quartic scalar couplings as well as the couplings of the scalar fields to the SM fermions and weak gauge bosons is given in Appendix~\ref{sec:a}.

It is worth noting that the CPV coupling $g_{123}$ is non-vanishing in the exact alignment limit $\epsilon\to 0$.\footnote{This is in sharp contrast with the C2HDM \cite{Low:2020iua}, where the corresponding   ${g}_{123}$ vanishes in the alignment limit.} We will see that this feature gives rise to a significant event rate for the triple Higgs boson final state.  Another trilinear coupling that does not vanish as $\epsilon\to 0$ is ${g}_{223}$, which will result in the quadruple Higgs final state.

\section{Experimental Constraints}\label{sec:3}

In this section, we study the viable parameter space of the CPVDM model, using empirical constraints coming from LHC Higgs measurements, electroweak oblique parameters, LHC direct searches of additional scalars, DM relic density, and DM direct and indirect search bounds.  Among input parameters in Eq.~(\ref{eq:input}), we take $m_{h_1}=125$~GeV and $v=246$~GeV.  We also remark in this section how the model is free from the electron EDM constraint up to at least two-loop level.

\subsection{LHC Higgs Measurements}\label{sec:3:1}

Due to the mixing between the doublet and singlet scalars, the 125-GeV Higgs boson may have a new invisible decay channel,  if $m_h \ge 2 m_\chi$, and its coupling strength to the SM fields is universally reduced.

\begin{figure}[t]
\centering
	\includegraphics[width=0.5\textwidth]{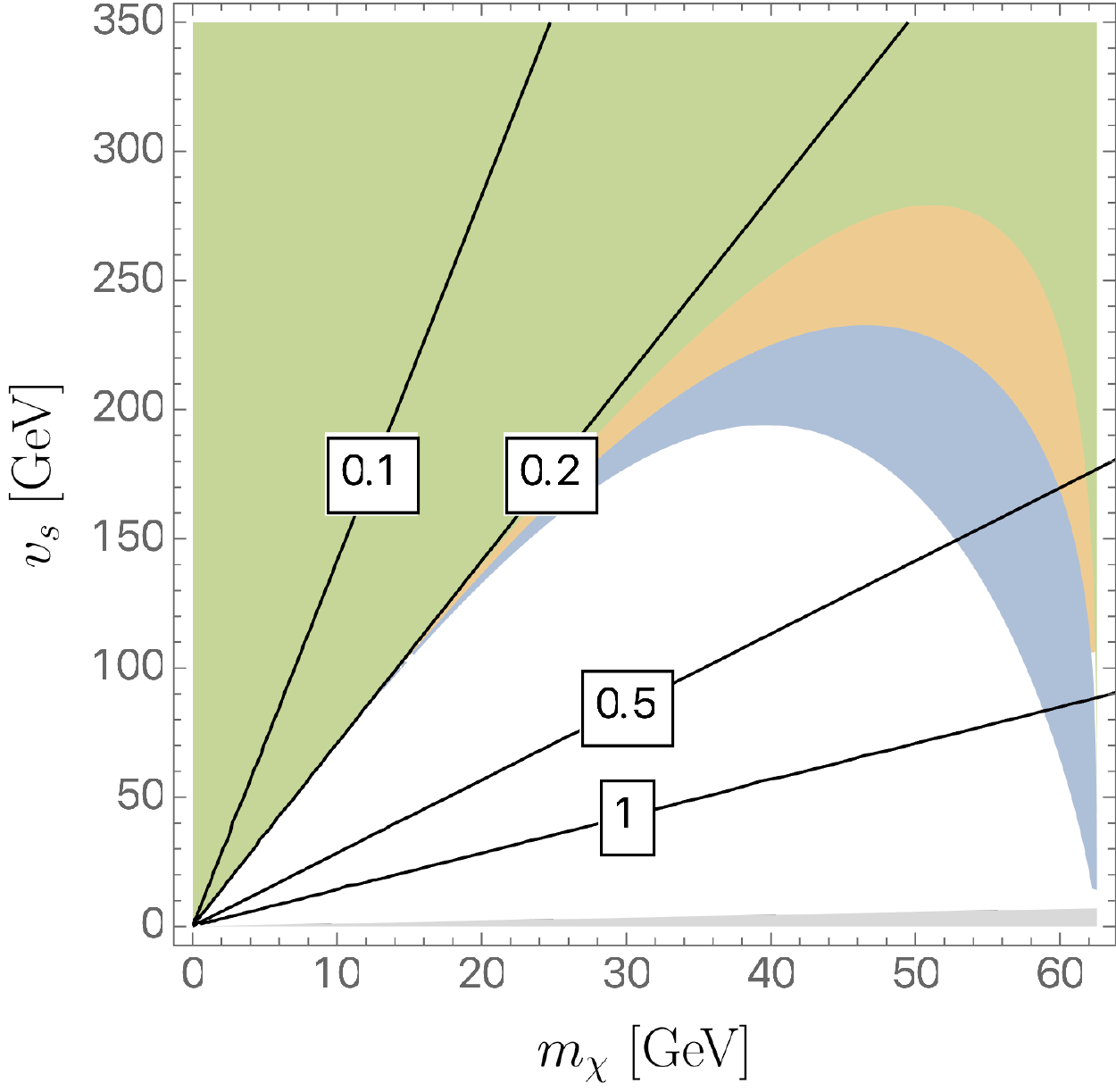}
\caption{\label{BR:XX} Constraints from the invisible decay of the 125-GeV Higgs in the $v_s$-$m_\chi$ plane with $\theta_{23}=0,\pi$ (green), $\pm\pi/4,\pm3\pi/4$ (orange), $\pm\pi/2$ (blue). The gray region at the bottom denotes the parameter space where $\lambda_\chi>4\pi$. Black lines are contours of $\lambda_\chi$.} 
\end{figure}

We first consider the constraint on the invisible decay rate given by CMS~\cite{CMS:2018yfx}:\footnote{The bound given by ATLAS~\cite{ATLAS:2019cid}, $BR(h_1\to{\rm inv}) < 0.26$ (95\% C.L.), is weaker than the one given by CMS.}
\begin{equation}
	BR(h_1\to{\rm inv}) < 0.19 ~({\rm 95\%~confidence~level~(C.L.)}) ~.
\end{equation}
For $m_\chi < m_{h_1}/2$, the $h_1\to\chi\overline{\chi}$ partial width is given by
\begin{equation}
	\Gamma(h_1\to\chi\overline{\chi}) = \frac{\epsilon^2}{8\pi}\frac{m_\chi^2}{v_s^2}m_{h_1}\Bigg(1-\frac{4m_\chi^2}{m_{h_1}^2}\Bigg)^{1/2}\Bigg[s_{23}^2\Bigg(1-\frac{4m_\chi^2}{m_{h_1}^2}\Bigg)+c_{23}^2\Bigg] 
	~. \label{BR:inv}
\end{equation}
In FIG.~\ref{BR:XX}, we show the constraint in the $v_s - m_\chi$ plane for $\epsilon=0.1$ and a few choices of $\theta_{23}$. The colored region is allowed by the invisible decay constraint. 
The small gray region { at the bottom of the plot denotes the region where $\lambda_\chi=\sqrt{2}m_\chi/v_s >4\pi$ and  violates the perturbativity bound}. The mixing angle $\theta_{23}$ has a significant impact on the constraints when $m_\chi$ is close to $m_{h_1}/2$, in which region the phase space suppression becomes prominent. If $h_1$ contains more CP-even component, {\it i.e.}, $s_{23}^2$ becomes larger, then the allowed phase space also becomes larger due to the extra factor of $1-4m_\chi^2/m_{h_1}^2$.

Next we consider the constraints coming from the measured Higgs signal strengths~\cite{Zyla:2020zbs} listed in TABLE.~\ref{Higgs:signal}.
\begin{table}[t]
\centering
\begin{tabular}{>{\centering\arraybackslash}p{2cm}|>{\centering\arraybackslash}p{4cm}}
\toprule
Channel & Signal Strength \\
\toprule
$ZZ$ & $\mu_{ZZ}=1.20\substack{+0.12 \\ -0.11}$ \\
\colrule
$W^+W^-$ & $\mu_{WW}=1.19\pm0.12$ \\
\colrule
$\gamma\gamma$ & $\mu_{\gamma\gamma}=1.11\substack{+0.10 \\ -0.09}$ \\
\colrule
$b\overline{b}$ & $\mu_{bb}=1.04\pm0.13$ \\
\colrule
$\tau^+\tau^-$ & $\mu_{\tau\tau}=1.15\substack{+0.16 \\ -0.15}$ \\
\colrule
$\mu^+\mu^-$ & $\mu_{\mu\mu}=0.6\pm0.8$ \\
\botrule
\end{tabular}
\caption{\label{Higgs:signal} Higgs signal strengths given in Ref.~\cite{Zyla:2020zbs}.}
\end{table}
In the model, the couplings of $h_1$ to the other SM fields are modified universally by a factor of $1-\epsilon^2/2$, leading to a reduction in the production rate by $1-\epsilon^2$. The branching ratios, however, remains the same unless new invisible decay channel opens up when $m_{h_1}> 2m_\chi$.  The strongest bound comes from the slightly enhanced signal strength in $\mu_{ZZ}$, which at {95\% C.L.} requires 
\begin{equation}
\label{inv:2sigma}
\vert\epsilon\vert< 0.125 \ .
\end{equation}
On the other hand, if $h_1 \to \chi\overline{\chi}$ is kinematically allowed, then the signal strengths of the SM channels are modified to be
\begin{equation}
\mu 
= (1-\epsilon^2)\big[1-BR(h_1\to\chi\overline{\chi})\big] = 1-\epsilon^2-\frac{\Gamma(h_1\to\chi\overline{\chi})}{\Gamma_{h_1}^{\rm SM}} + \mathcal{O}(\epsilon^4) ~ ,
\end{equation}
where $\Gamma_{h_1}^{\rm SM}$ is the total decay width of the 125-GeV Higgs predicted by the SM. Seeing that such modifications would make the constraint on $\epsilon$ even stronger, we do not consider this case and only explore the case of $m_\chi>m_{h_1}/2$ in the benchmark studies, where we fix $\epsilon=0.1$.

\subsection{Electroweak Oblique Corrections}\label{sec:3:2}

\begin{figure}[t]
\centering
	\includegraphics[width=0.5\textwidth]{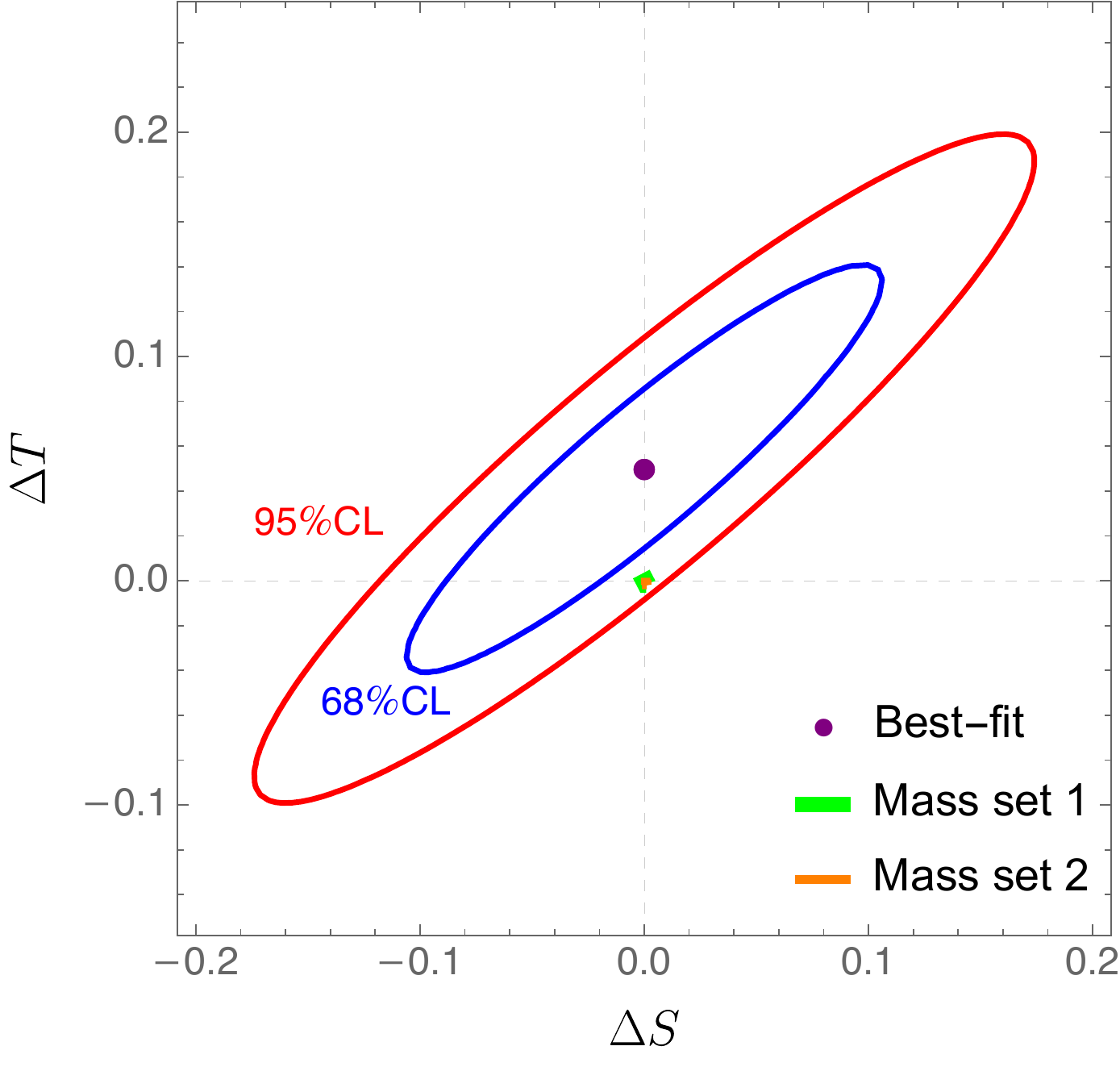}
\caption{\label{EWPO} {Oblique corrections for $\epsilon=0.1$ and the two choices of masses in Eq.~(\ref{eq:benchmark_masses}). We remark that mass set 2 (orange) is on top of mass set 1 (green). }} 
\end{figure}

We now consider the Peskin-Takeuchi $S$ and $T$ parameters defined in Ref.~\cite{Peskin:1991sw}. The current fits given by PDG~\cite{Zyla:2020zbs} are
\begin{equation}
\begin{aligned}
	\Delta S &= 0.00\pm0.07 ~, \\
	\Delta T &= 0.05\pm0.06 ~,
\end{aligned}
\end{equation}
with a correlation of $0.92$. In our model, {because the vector bosons only couple to the physical scalars through their $\phi_1$-components, the mixing of which is determined entirely by $\epsilon$ and $\theta_{12}$ as shown in Eq.~(\ref{eq:mixing})\footnote{$\theta_{23}$ only parametrizes the mixing between $\phi_2$ and $a$, but not the gauge couplings, and hence does not take part in the oblique corrections.},} $S$ and $T$ parameters only depend on $m_{h_2},m_{h_3},\epsilon$ and, to a much less extent, $\theta_{12}$. We fix $\epsilon=0.1$ as in Sec.~\ref{sec:3:1}, and choose two sets of heavy scalar masses:
\begin{align}
\begin{split}
(m_{h_2},m_{h_3}) = (280,420)~\mbox{GeV}
~, \\
(m_{h_2},m_{h_3}) = (280,600)~\mbox{GeV}
~.
\end{split}
\label{eq:benchmark_masses}
\end{align}
The first set is chosen to allow the $h_3\to h_1h_2$, $h_2\to2h_1$, and $h_3\to2h_1$ decays, while the second further allows the $h_3\to2h_2$ decay. 
For both mass sets, the above constraint can be satisfied within ${1-2}\sigma$ for all possible values of $\theta_{12}$. In fact the oblique corrections have very little dependence on $\theta_{12}$, whose contributions are suppressed by $\epsilon^2$.  We show the 68\% C.L. and 95\% C.L. contours in the $\Delta T$-$\Delta S$ plane, as well as the values for both sets of masses, in FIG.~\ref{EWPO}.  The detailed formulas for the electroweak oblique observables are given in Appendix~\ref{sec:b}.

\subsection{LHC  Searches for Heavy Scalars}\label{sec:3:3}

Here we consider constraints from direct searches of heavy neutral scalars at the LHC, focusing on the diboson final states: $WW$, $ZZ$ and $h_1h_1$. The $t\bar{t}$ channel is less stringent. The light decay channels such as $b\bar{b}$, $\tau^+\tau^-$, and $\gamma\gamma$ are also less stringent because of suppressed decay BRs. \footnote{Unlike in the C2HDM, we do not have to consider $h_i\to h_1Z$, $i=2,3$, decays, which are absent in our model because the singlet scalar does not contain any ``eaten'' Goldstone bosons. }

Because the singlet scalar $S$ does not couple to the SM gauge bosons and fermions directly, $h_2$ and $h_3$ couple to the SM gauge bosons and fermions only through their $\phi_1$ component. As such, their productions will go through the gluon-fusion (ggF) channel and are suppressed by the alignment parameter $\epsilon$:
\begin{align}
 \label{eq:direct} 
\begin{split}
& \sigma(gg\to h_i)=R_{i1}^2\ \sigma^{\rm SM}(gg\to h_i)
~,\\
& \Gamma(h_i\to f_{\rm SM})=R_{i1}^2\ \Gamma^{\rm SM}(h_i\to f_{\rm SM})
~,
\end{split}
\end{align}
where $R_{11}=1-\epsilon^2/2$, $R_{21}=-\epsilon s_{12}$, and $R_{31}=-\epsilon c_{12}$. In addition, $\sigma^{\rm SM}(gg\to h_i)$ and $\Gamma^{\rm SM}(h_i\to f_{\rm SM})$ denote the SM production rate and decay partial width at the mass $m_{h_i}$, which we obtain from Ref.~\cite{LHCHiggsCrossSectionWorkingGroup:2016ypw}.
In addition to direct two-body decays from $h_i\to VV/h_1h_1$, we also include Higgs-to-Higgs decays such as 
$h_3\to (h_2\to 2h_1)+h_1$.

We base our constraints on those in Refs.~\cite{ATLAS:2018ili,ATLAS:2018hqk,CMS:2017rpp,ATLAS:2018fpd,CMS:2017hea,ATLAS:2018uni,ATLAS:2018dpp,CMS:2018tla,ATLAS:2018rnh,CMS:2018qmt,CMS:2018ipl,ATLAS:2019qdc,CMS:2018amk,CMS:2019bnu,ATLAS:2020fry,ATLAS:2020tlo}.  As seen in Eq.~(\ref{eq:direct}), the direct search constraints are all sensitive to $\vert\epsilon\vert$, which suppresses the production rates by a factor of $\epsilon^2$. Moreover, while the $WW$ and $ZZ$ constraints only depend on $\epsilon$ and $\theta_{12}$ in addition to $m_{h_2}$ and $m_{h_3}$, the $h_1h_1$ constraints further depend on $\theta_{23}$, $\delta_2$, and $\delta_3$ through their participation in the $g_{112}$ and $g_{113}$ couplings. Moreover, $\theta_{23}$ always shows up in $g_{112}$ and $g_{113}$ in the combination of  $\theta_{\delta_3}-(\theta_{12}+2\theta_{23})$, as can be seen in Eq.~(\ref{eq:gijkcoupling}).  We choose to fix $\vert\delta_3\vert=3.5$ and focus on the effects of $\delta_2$ on the constraints.   In order to maximize the $3h_1$ cross sections, we further maximize $\vert{g}_{123}\vert$ by choosing $\theta_{\delta_3}-2(\theta_{12}+\theta_{23})=-\pi/2$.  Finally, we choose $\theta_{12}$ in a way that the ggF production rates of $h_2$ and $h_3$ are similar, which implies $\theta_{12}=0.73$ for the first mass set and $\theta_{12}=0.41$ for the second mass set  in Eq.~(\ref{eq:benchmark_masses}).

\begin{figure}[t]
\centering
	\includegraphics[width=0.4\textwidth]{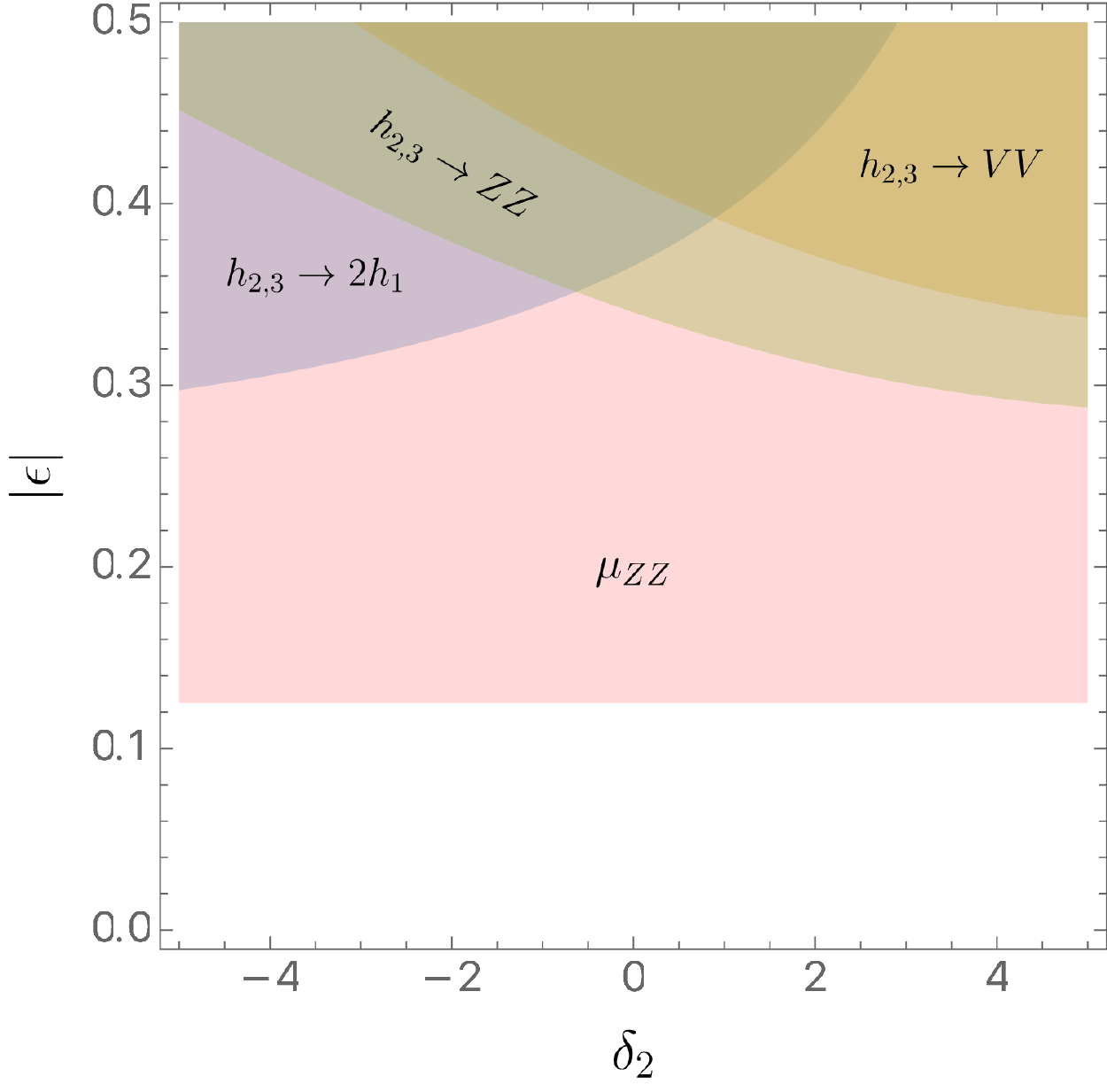}
	\includegraphics[width=0.4\textwidth]{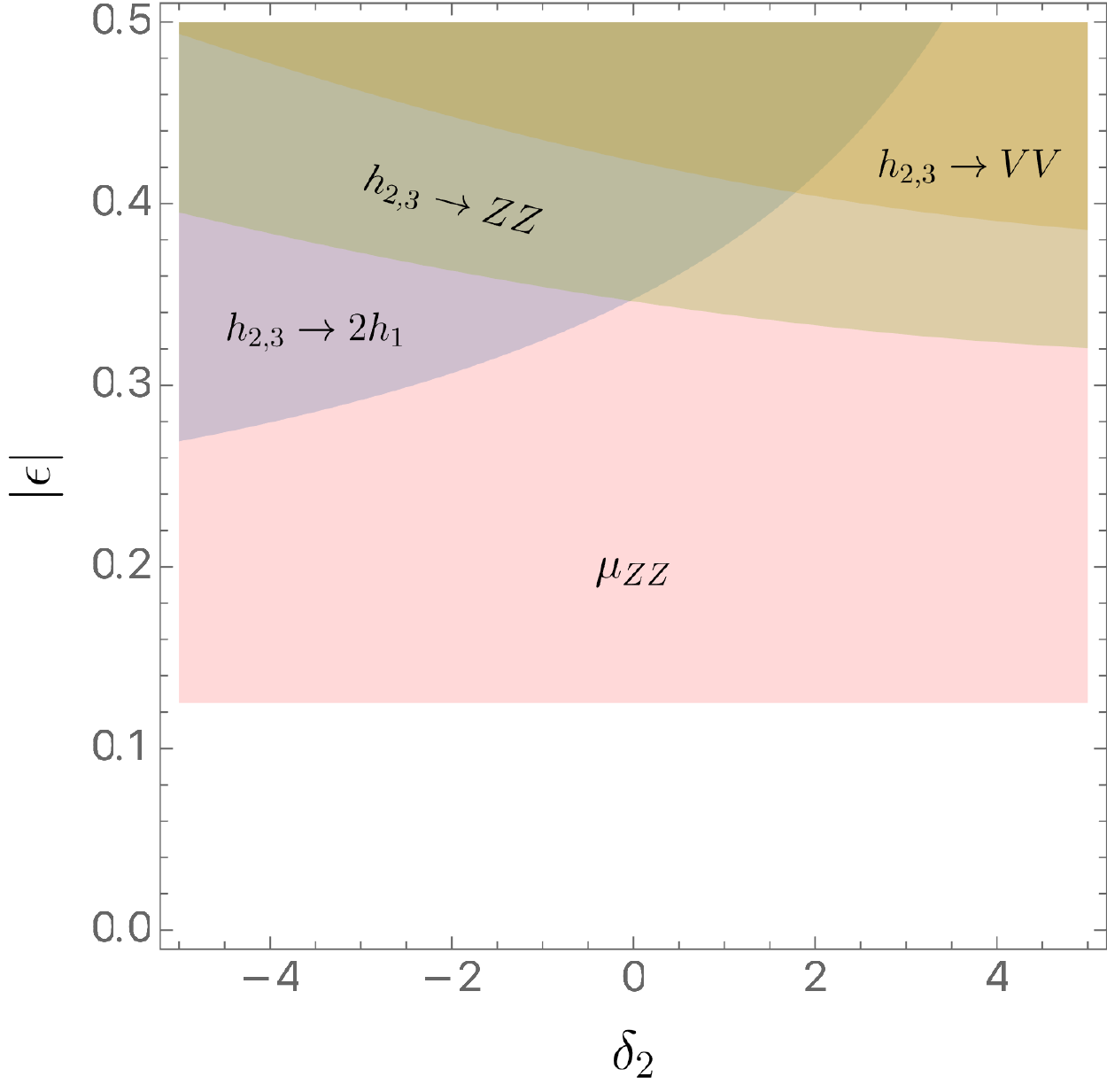}
\\
\vspace{-0.4cm}
(a) \hspace{6cm} (b)
\\
\vspace{0.5cm}
	\includegraphics[width=0.4\textwidth]{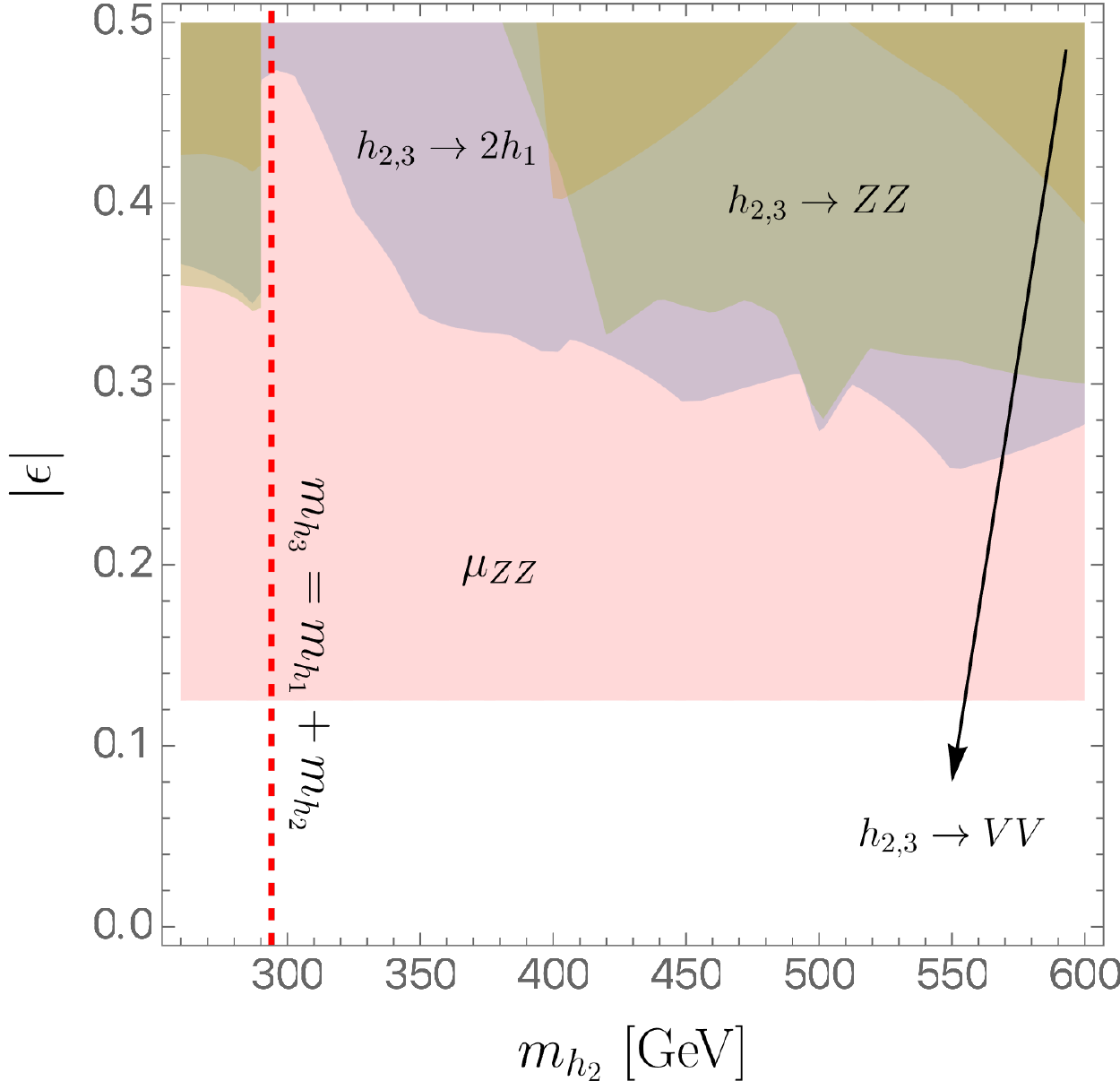}
	\includegraphics[width=0.4\textwidth]{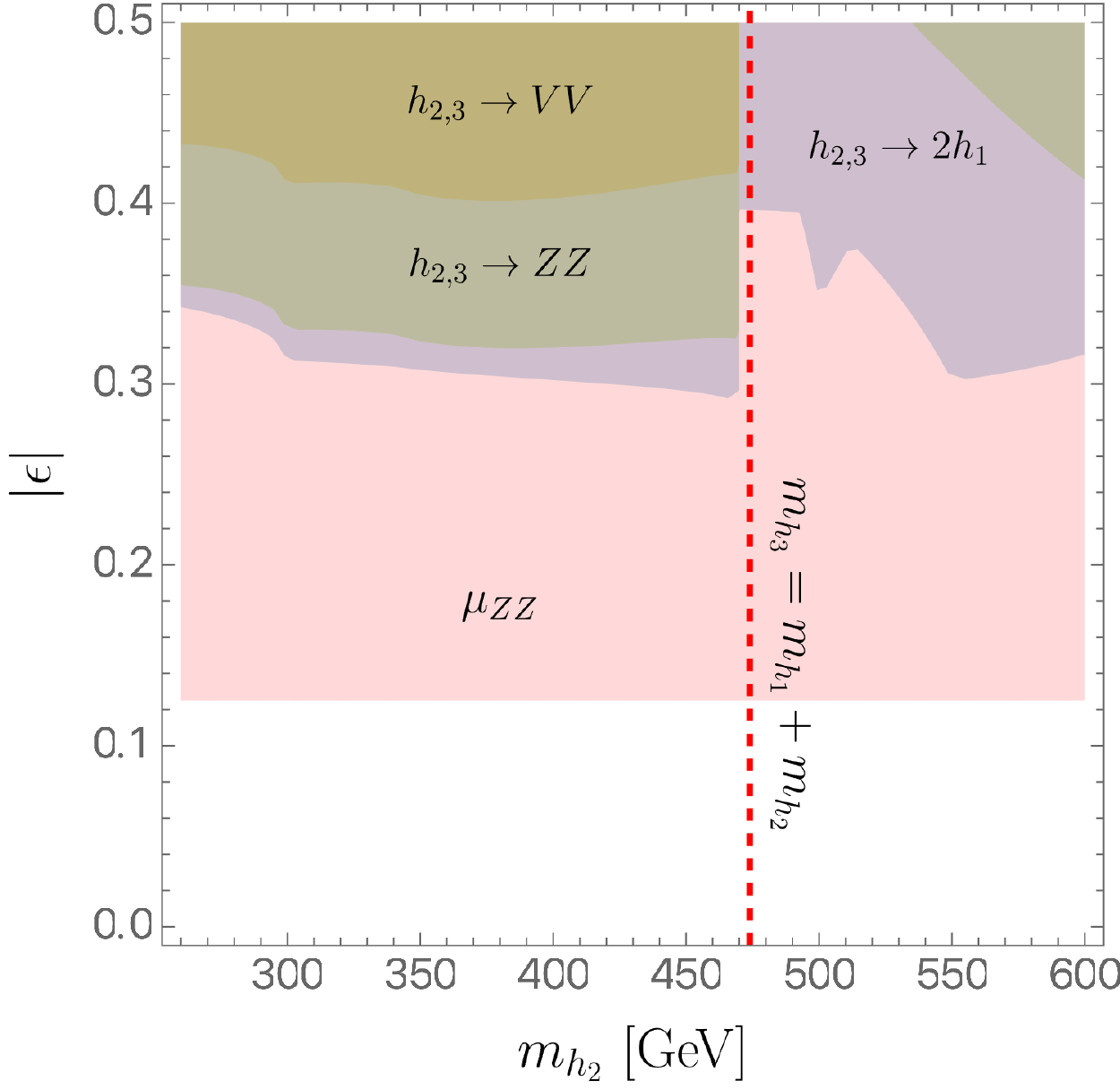}
\\
\vspace{-0.4cm}
(c) \hspace{6cm} (d)
\caption{\label{LHC:direct} Parameter space excluded by the LHC direct search constraints in the $\delta_2$-$\vert\epsilon\vert$ plane for (a) mass set 1 and (b) mass set 2, and in the $m_{h_2}$-$\vert\epsilon\vert$ plane with (c) $(\delta_2,m_{h_3})$ $=$ $(-0.5,420~{\rm GeV})$ and (d) $(\delta_2,m_{h_3})$ $=$ $(-0.5,600~{\rm GeV})$. The parameter space excluded by the $\mu_{ZZ}$ measurement at 95\% C.L. is also included in the plots.  The vertical red dashed lines mark the onset of the $h_3 \to h_1 + h_2$ decay.}
\end{figure}

We examine the direct search constraints in the $\delta_2$-$\vert\epsilon\vert$ plane for the two mass sets, and in the $m_{h_2}$-$\vert\epsilon\vert$ plane with $(\delta_2,m_{h_3})=(-0.5,420~{\rm GeV})$ and $(\delta_2,m_{h_3})=(-0.5,600~{\rm GeV})$, respectively, focusing on the region where $\delta_2\in[-5,5]$, $m_{h_2}\in[260,600]$~GeV, and $\vert\epsilon\vert\in[0,0.5]$.  For definiteness as well as to impose the constraints on the parameter space in the strictest manner, we also neglect the $h_{2,3}\to\chi\bar{\chi}$ decays. We present the results in FIG.~\ref{LHC:direct}, where we also include the $\mu_{ZZ}$ constraint at 95\% C.L. in Eq.~(\ref{inv:2sigma}). Note that in FIGS.~\ref{LHC:direct}(c) and (d), there are two sudden jumps caused by the onset of the $h_3\to h_2h_1$ decay when $m_{h_3}\ge m_{h_1}+m_{h_2}$. It can be seen that the direct search constraints are all less stringent than the $\mu_{ZZ}$ constraint, and thus for our choice of $\epsilon=0.1$, both mass sets are completely safe from the LHC direct search constraints within the specified region.

\subsection{Comments on electron EDM Constraints}\label{sec:3:4.1}

We remark in this section that our CPVDM model does not generate new EDM contributions up to at least two-loop level, in sharp contrast with the C2HMDs. This is mainly due to the fact that new sources of CPV in our model are confined to the scalar and dark sectors:  the CPV interactions take place among the scalars, or between $\chi\overline{\chi}$ and the messenger scalar $S$. There is no new source of CPV in the visible fermionic sector as the singlet scalar $S$ does  not have Yukawa interactions with the SM fermions. Therefore, even though the 125-GeV SM-like Higgs could have a component in $S$ due to the mass mixing, such a component does not introduce any CP-odd coupling of the 125-GeV Higgs to the SM fermions. Similar consideration applies to the heavy scalars couplings to SM fermions, which are induced only through the doublet component and remain CP-even.  This explains why no electron EDM appears at one loop.

\begin{figure}[t!]
\centering
	\includegraphics[width=0.40\textwidth]{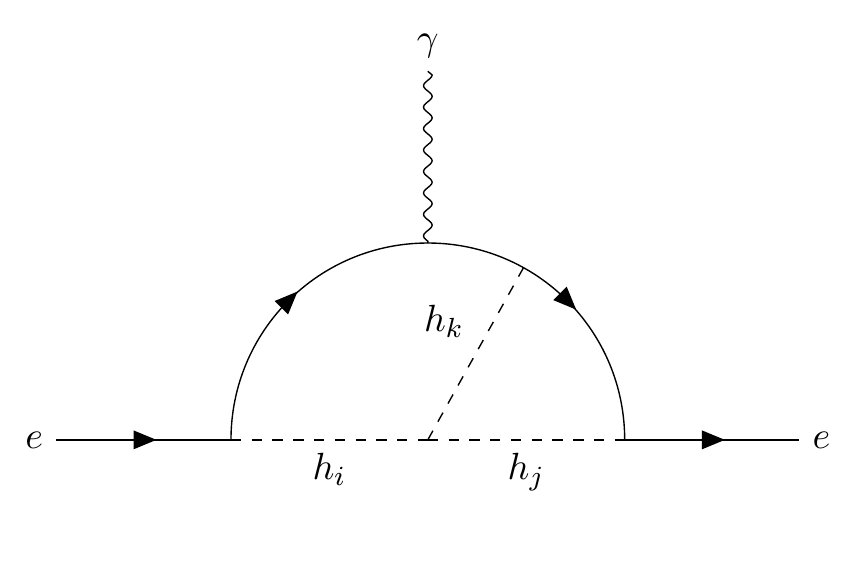}
	\includegraphics[width=0.40\textwidth]{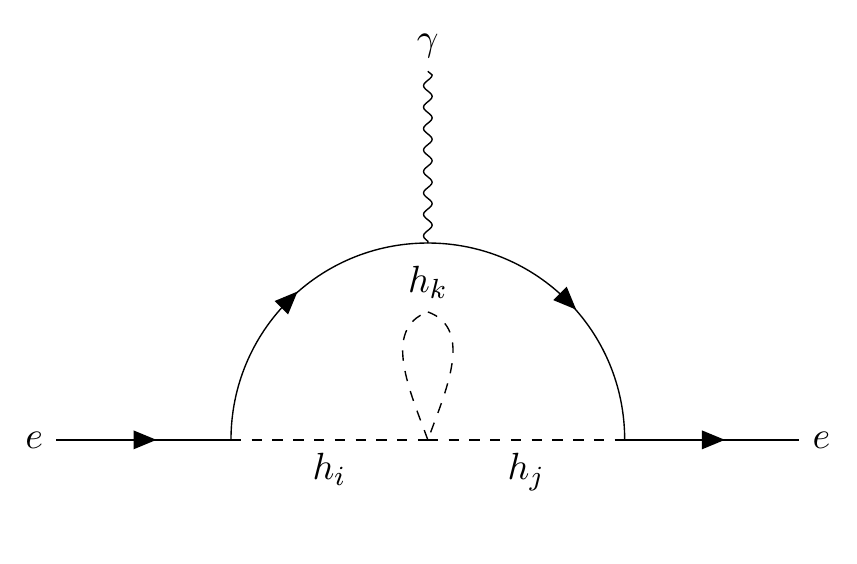}
\\
\vspace{-0.4cm}
(a) \hspace{6cm} (b)
\\
\vspace{0.5cm}
	\includegraphics[width=0.40\textwidth]{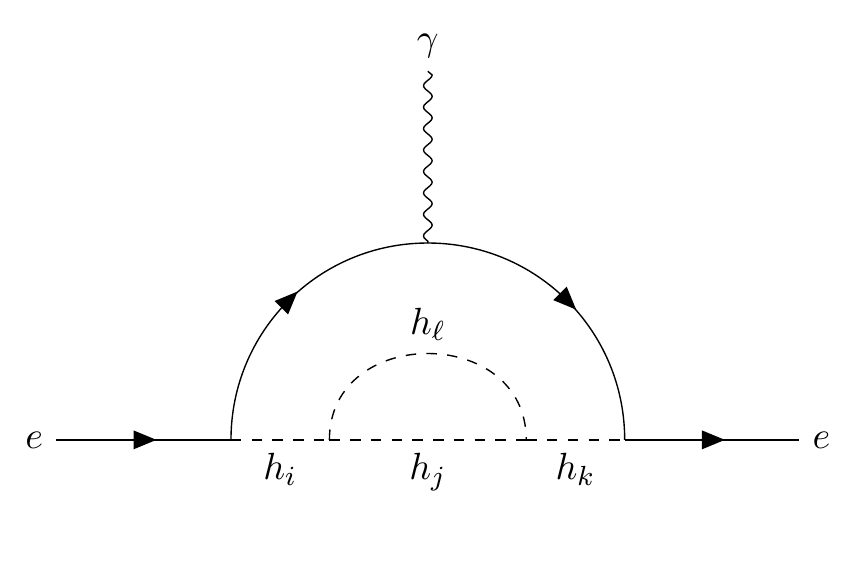}
	\includegraphics[width=0.40\textwidth]{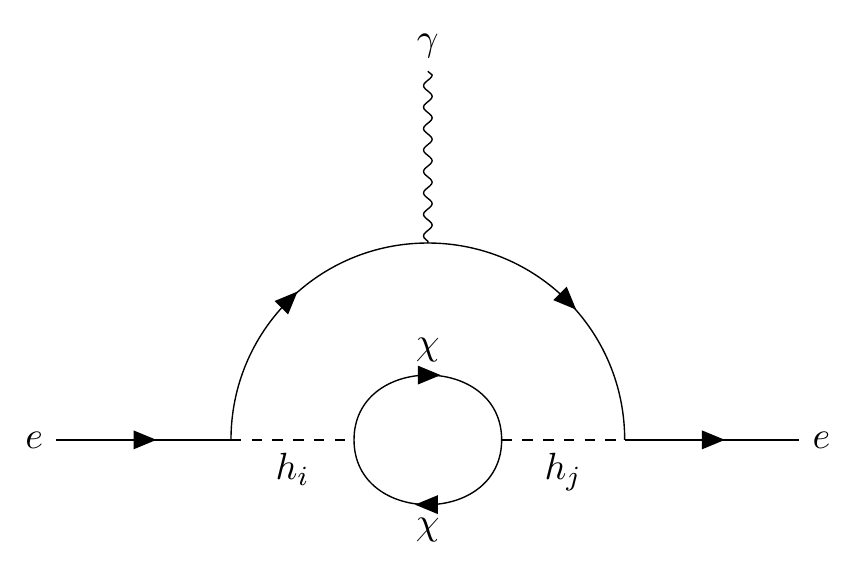}
\\
\vspace{-0.4cm}
(c) \hspace{6cm} (d)
\caption{\label{EDM:Feyn} {Two-loop CPV diagrams that include (a) three scalar lines that all attach to the electron lines, (b) three scalar lines with an internal loop, (c) four scalar lines with an internal scalar loop, and (d) two scalar lines along with a $\chi$-loop.}}
\end{figure}

Now we exhibit potential two-loop contributions to $e-e-\gamma$ coupling in our model in FIG.~\ref{EDM:Feyn}.  To introduce CPV into the three-point electron-photon interaction, we must insert at least one trilinear/quartic CPV scalar vertex or one CPV scalar-$\chi\bar{\chi}$ vertex into the internal loops.  In the first case, the scalars have to either all attach to the electron lines, such as that in FIG.~\ref{EDM:Feyn}(a), or form an internal loop, such as that in FIGS.~\ref{EDM:Feyn}(b) and (c).  As for the second case, since $\chi$ only interacts with the scalars, it must form an internal loop, as shown in FIG.~\ref{EDM:Feyn}(d).  For the cases of FIGS.~\ref{EDM:Feyn}(a), (b), and (c), no factor of $\gamma^5$ would appear, and hence no EDM would be induced.  As to FIG.~\ref{EDM:Feyn}(d), after taking the Dirac trace of the fermion loop, no Lorentz-invariant terms of the form $\epsilon^{\alpha\beta\rho\sigma}p_{1\alpha}p_{2\beta}p_{3\rho}p_{4\sigma}$ would be induced, where $\epsilon^{\alpha\beta\mu\nu}$ is the rank-four Levi-Civita symbol and $p_{1,2,3,4}$ are generic four-momenta, and hence there is no EDM contribution either.

\subsection{Muon Anomalous Magnetic Dipole Moment}\label{sec:3:4.2}

\begin{figure}[t!]
\centering
	\includegraphics[width=0.40\textwidth]{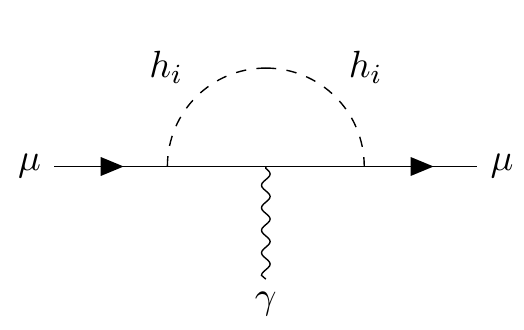}
	\includegraphics[width=0.40\textwidth]{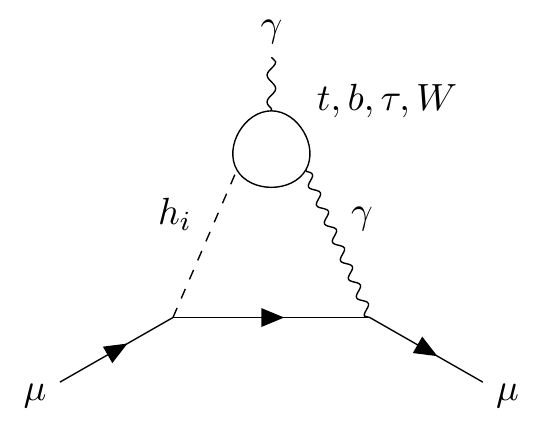}
\\
\vspace{-0.4cm}
(a) \hspace{6cm} (b)
\caption{\label{muon} (a) One-loop and (b) two-loop Barr-Zee diagrams contributing to $(g-2)_\mu$ in our model. }
\end{figure}

In this section, we briefly comment on the contributions to the muon anomalous magnetic dipole moment, $(g-2)_\mu$, in our model. The latest measurement was made in the E989 experiment at Fermilab, and the result was given by~\cite{Muong-2:2021ojo}
\begin{equation}\label{mu:exp}
	a_\mu^{\rm FNAL} = 116592040(54)\times10^{-11} ~,
\end{equation}
while the SM prediction is given by~\cite{Aoyama:2020ynm}
\begin{equation}\label{mu:SM}
	a_\mu^{\rm SM} = 116591810(43)\times10^{-11} ~,
\end{equation}
leading to a 4.2$\sigma$ discrepancy
\begin{equation}\label{mu:diff}
	\Delta a_\mu = (251\pm59)\times10^{-11} ~.
\end{equation}
The leading contributions are the one-loop diagrams and the two-loop Barr-Zee diagrams with top-, bottom-, $\tau$-, and $W$-loops running in the loop, as demonstrated in FIG.~\ref{muon}. Denoting their contributions by $\Delta a_\mu^{(1)}$ and $\Delta a_\mu^{(2)}$, respectively, we have~\cite{Ilisie:2015tra,Chen:2020tfr,Chiang:2021pma,Chen:2021jok}
\begin{equation}
	\Delta a_{\mu}^{(1)} = \sum_i\epsilon^2\kappa_i\frac{m_\mu^2}{8\pi^2v^2}\int_0^1dx\frac{\tau^\mu_ix^2(2-x)}{1-x(1-\tau^\mu_ix)} ~,
\end{equation}
\begin{equation}
\begin{aligned}
	\Delta a_{\mu}^{(2)} &= \sum_i\epsilon^2\kappa_i\frac{\alpha m_\mu^2}{8\pi^3v^2}\Bigg[ \sum_{f=t,b,\tau}N_c^fQ_f^2\int_0^1dx\tau^f_i\frac{2x(1-x)-1}{\tau^f_i-x(1-x)}\ln\left(\frac{\tau^f_i}{x(1-x)}\right) \\
	&\quad +\frac{1}{2}\int_0^1dx\frac{x\left[3x(4x-1)+10\right]\lambda_i-x(1-x)}{\lambda_i-x(1-x)}\ln\left(\frac{\lambda_i}{x(1-x)}\right) \Bigg] ~,
\end{aligned}
\end{equation}
where $\tau^f_i=m_f^2/m_{h_i}^2$, $\lambda_i=m_W^2/m_{h_i}^2$, and
\begin{equation}
	\kappa_1=-1, ~\kappa_2=s_{12}^2, ~\kappa_3=c_{12}^2 ~.
\end{equation}
Note that they are both suppressed by $\epsilon^2$.  With the two chosen scalar mass sets, we have $\Delta a_\mu\sim\mathcal{O}\left(10^{-13}\right)$, which is negligible compared to Eq.~(\ref{mu:diff}). Thus, $(g-2)_\mu$ cannot be addressed in our model.

\subsection{DM Constraints: Relic Density, Direct and Indirect Searches}\label{sec:3:4}

\begin{figure}[t]
\centering
	\includegraphics[width=0.4\textwidth]{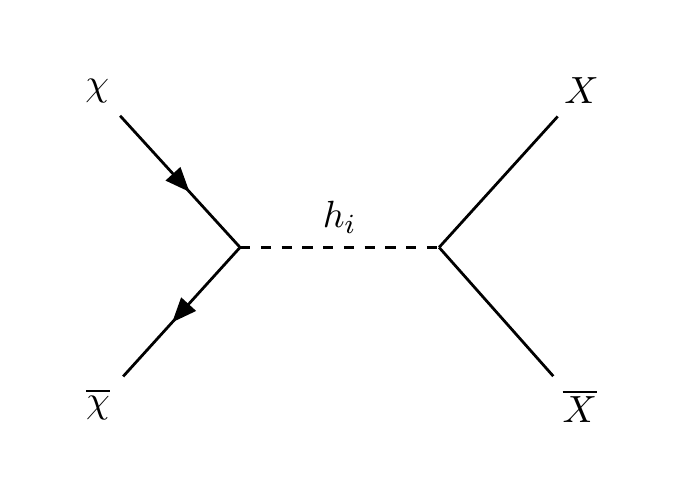}
\caption{\label{DMannih} DM annihilation processes. $h_i=h_1,h_2,h_3$, $X=q,\ell^-,g,W^+,Z,h_j$, and $\overline{X}=\overline{q},\ell^+,g,W^-,Z,h_k$.}
\end{figure}

\begin{figure}[t]
	\includegraphics[width=0.4\textwidth]{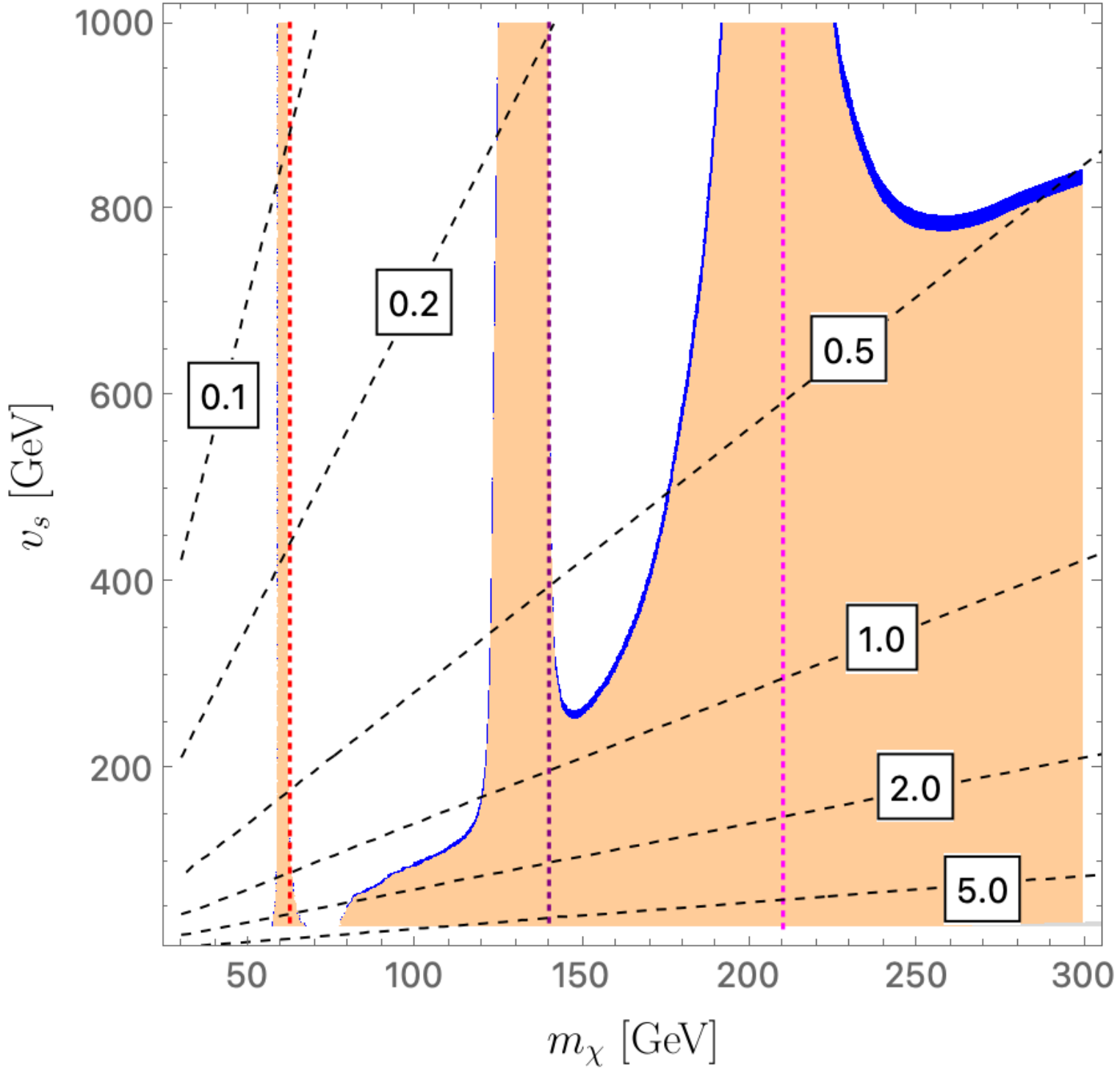}
	\includegraphics[width=0.4\textwidth]{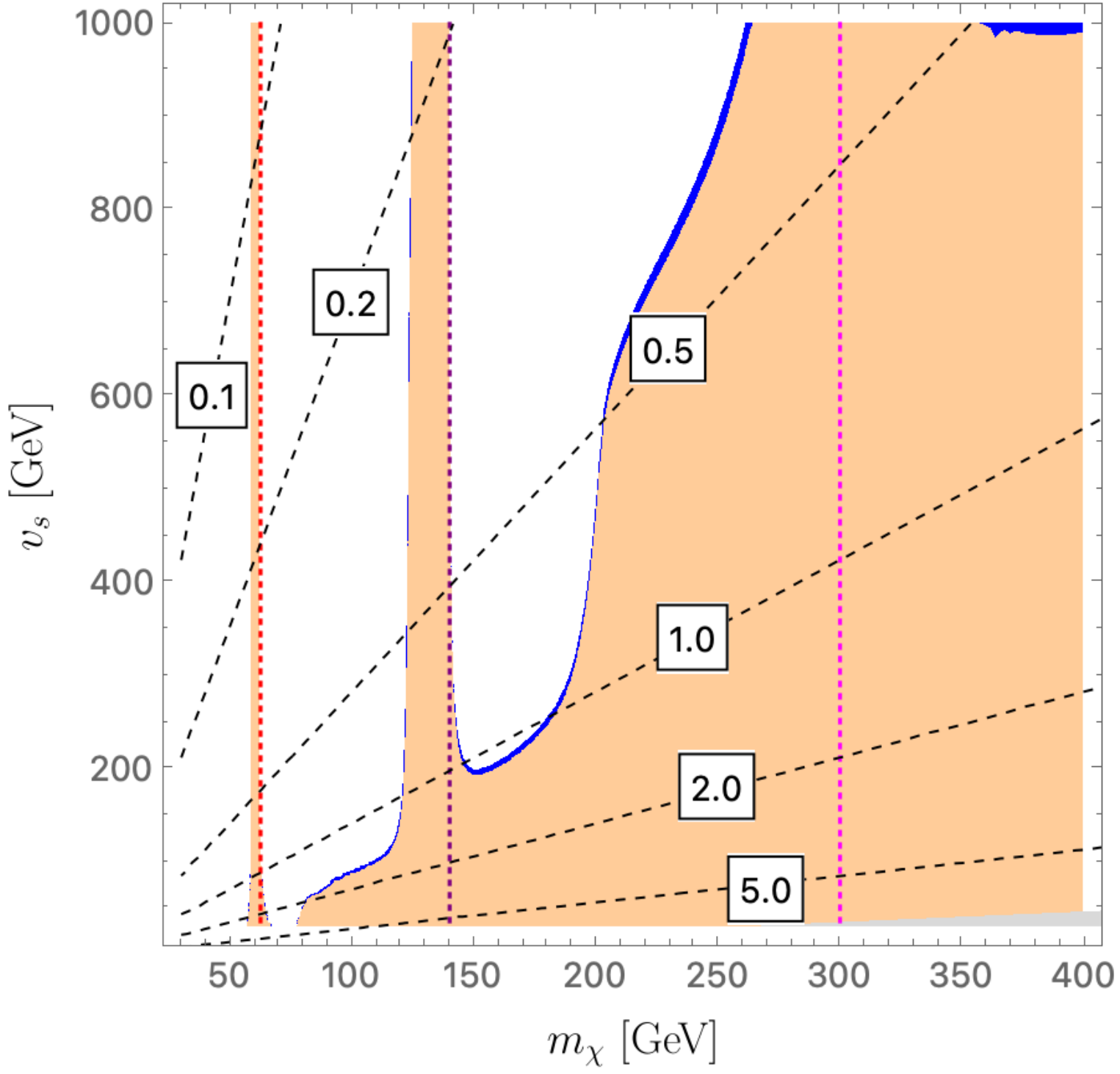}
	\\
	\vspace{-0.4cm}
(a) \hspace{6cm} (b)
\caption{\label{DMrelic}  DM relic density constraint in the $v_s$~vs.~$m_\chi$ plane for (a) mass set 1 and (b) mass set 2, assuming $\delta_2=-0.5$. The blue regions denote the parameter space that falls within the experimental 2$\sigma$ bounds, and the orange regions those below the lower 2$\sigma$ bound. The red, purple, and magenta dotted line denote respectively the contour where $m_\chi=m_{h_1}/2$, $m_{h_2}/2$, and $m_\chi=m_{h_3}/2$. {Black dashed lines are contours of $\lambda_\chi$.}}
\end{figure}

\begin{figure}[t]
	\includegraphics[width=0.4\textwidth]{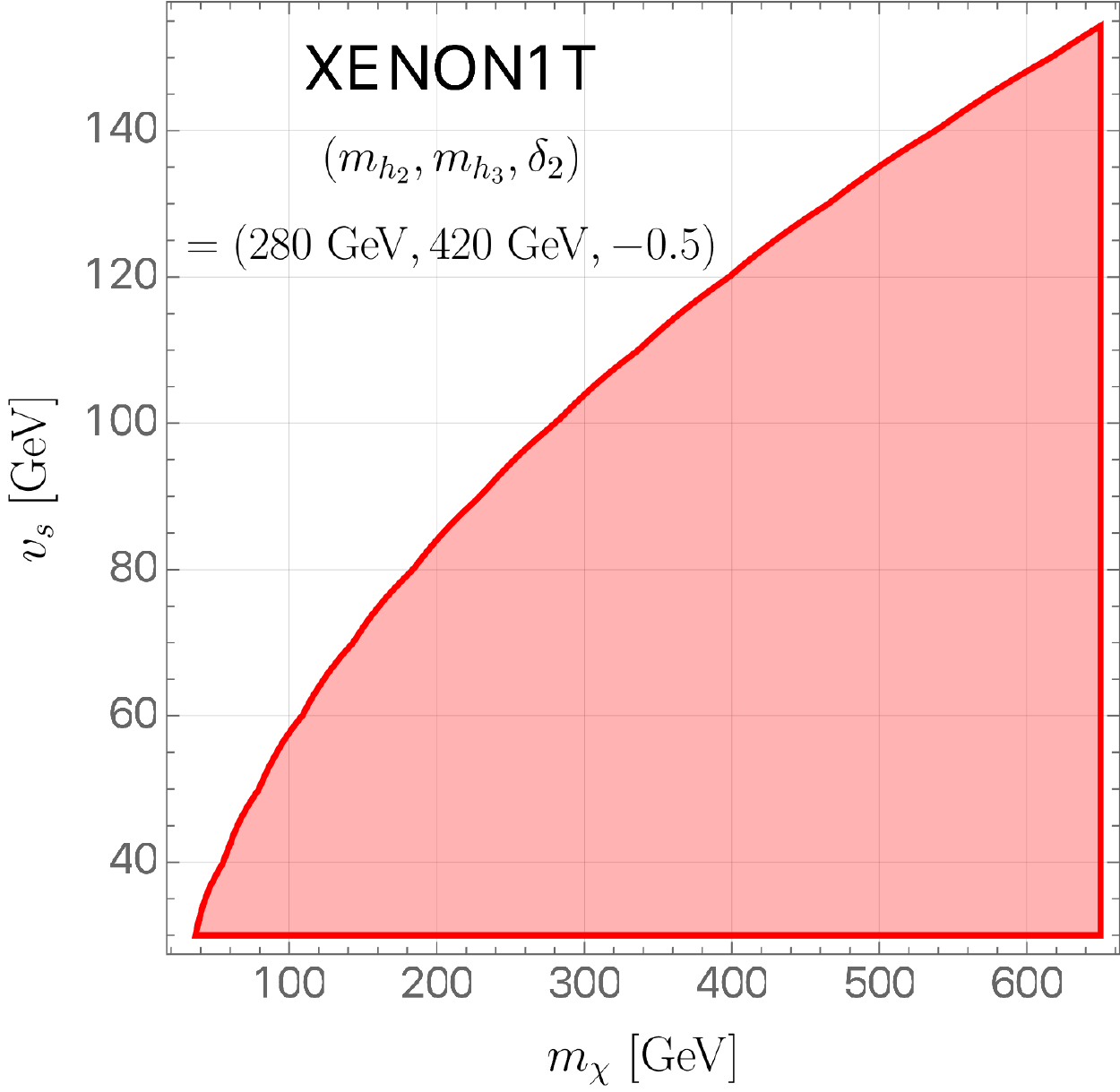}
	\includegraphics[width=0.4\textwidth]{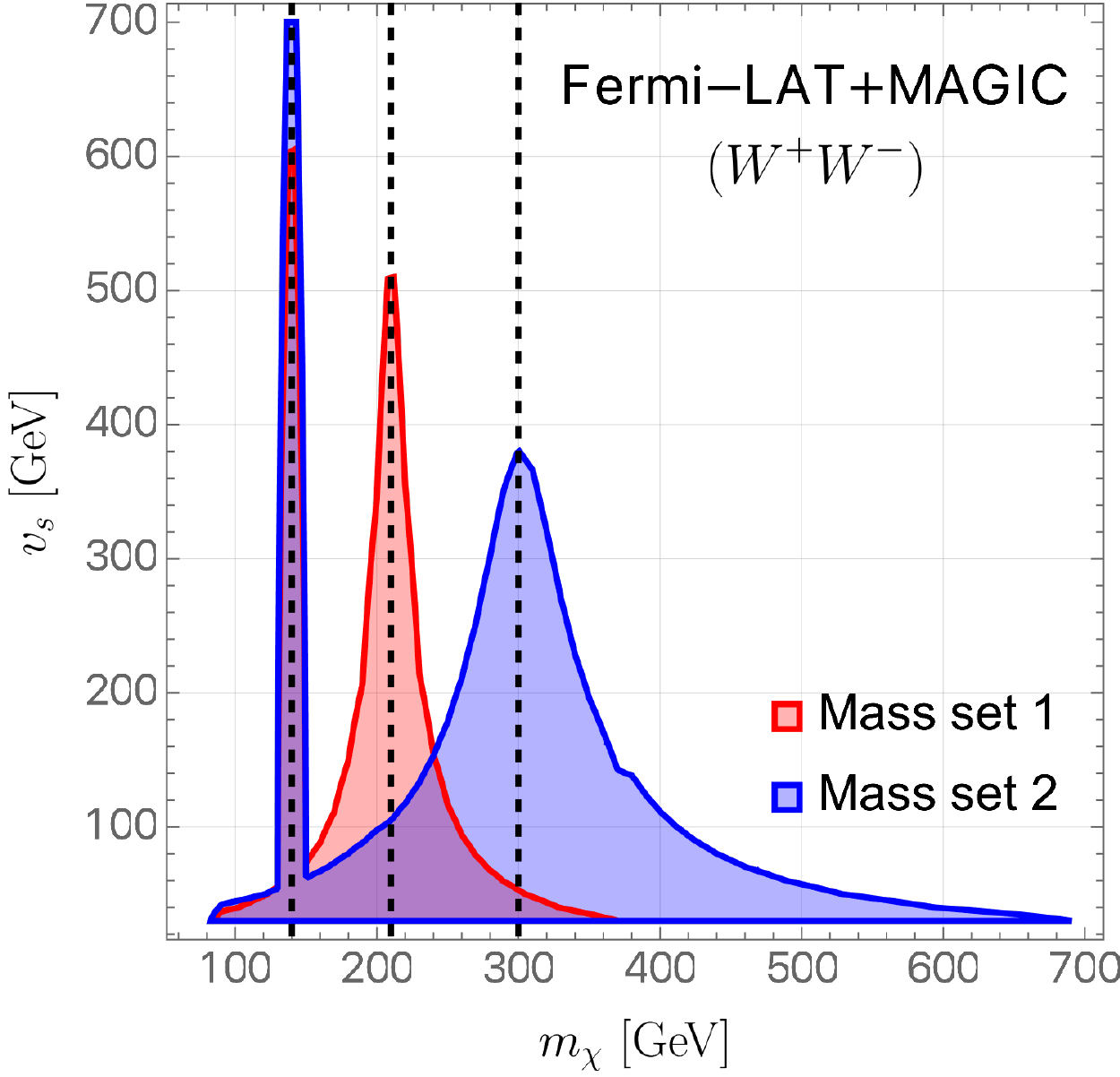}
\\
\vspace{-0.4cm}
(a) \hspace{6cm} (b)
\caption{\label{DM:Detect} (a) The 90\% C.L. excluded parameter space from XENON1T on the spin-independent proton-neutron-averaged DM direct detection cross sections in the $m_\chi$-$v_s$ plane for mass set 1 in Eq.~(\ref{eq:benchmark_masses}) with $\delta_2=-0.5$. The bound is similar for the mass set 2 and is not sensitive to the value of $\delta_2$. (b) The 95\% C.L. excluded parameter space for the $W^+W^-$ channel from the Fermi-LAT+MAGIC combined analysis in the $m_\chi$-$v_s$ plane for the DM pair annihilation rate
for the two scalar mass sets.  We also mark the $h_2,h_{3}^{\rm mass~set~1},h_{3}^{\rm mass~set~2}$ resonances near $m_\chi=140,210,300$~GeV (black dashed lines).}
\end{figure}

The DM relic density is measured to be $\Omega_\chi h^2=0.1197\pm0.0022$ \cite{Planck:2015fie}. The $\chi\overline{\chi}$ annihilation processes are shown in FIG.~\ref{DMannih}.
We use \texttt{micrOMEGAs}~\cite{Belanger:2004yn,Belanger:2006is,Belanger:2008sj,Belanger:2010pz,Belanger:2013oya} to calculate the relic density of $\chi$, which is mainly determined by $m_\chi=\lambda_\chi v_s/\sqrt{2}$ and $v_s$.  Other input parameters such as $\delta_2$ do not have a significant impact on the constraints.  In FIG.~\ref{DMrelic} we show the relic density constraints in the $v_s$-$m_\chi$ plane for {the two mass sets mentioned in Eq.~(\ref{eq:benchmark_masses})} with $\delta_2=-0.5$.  The small blue regions denote the parameter space that has a relic density within the experimental 2$\sigma$ bounds, and the orange regions those below the lower 2$\sigma$ bound. We also plot the $m_\chi=m_{h_1}/2$ contour (red dotted), the $m_\chi=m_{h_2}/2$ contour (purple dotted), and the $m_\chi=m_{h_3}/2$ contour (magenta dotted) to show the resonance effect.

As can be seen in FIG.~\ref{DMrelic}, the annihilation process is quite efficient for a DM mass that is sufficiently heavy, $m_\chi \agt 100$ GeV, and/or small $v_s$. In particular, a small $v_s$ increases $\lambda_\chi$ for a fixed $m_\chi$, which makes the annihilation rate larger. When the DM mass is close to the resonance region, $m_\chi \approx m_{h_i}/2$, $i=1,2, 3$, {the annihilation rate becomes enhanced and the relic density reduced, as can be seen from the plot. For our benchmark study, we choose the following four DM masses: $m_\chi=156,187,280,420$~GeV, which will be further studied in Sec.~\ref{sec:4}.

As for the DM direct searches, we quote the results of XENON1T~\cite{XENON:2018voc}.  Since $\delta_2$ is only relevant to scalar interactions, it does not have a significant impact on the DM-nucleon scattering at the leading order. Furthermore, both scalar mass sets in Eq.~(\ref{eq:benchmark_masses}) give roughly the same results.  In FIG.~\ref{DM:Detect}(a), we show the experimental constraint in the $m_\chi$-$v_s$ plane for mass set 1 and $\delta_2=-0.5$, taking the average of the spin-independent DM-proton and -neutron scattering cross sections.

Finally, for indirect DM searches, we quote the results of the Fermi-LAT+MAGIC combined analysis~\cite{MAGIC:2016xys} to constrain dark matter annihilation in the $b\bar{b}$ and $W^+W^-$ channels. In  the $2h_1$ channel we use the recent results given by MAGIC~\cite{MAGIC:2021mog}.  We perform a scan over $m_\chi,v_s,\delta_2$ for the $\chi$-pair annihilation rates into different final states. The dominant annihilation channel is largely determined by kinematics: for $m_\chi\lesssim85$~GeV, $\chi$-pairs mainly annihilate into $b\bar{b}$; for $85~{\rm GeV}\lesssim$~$m_\chi\lesssim200$~GeV, they  mainly annihilate into $W^+W^-$ or $h_1h_1$; for $m_\chi\gtrsim200$~GeV, they mainly annihilate into heavy scalars, and occasionally to $W^+W^-$ or $h_1h_1$.

Again $\delta_2$ does not have a considerable impact for annihilations into the SM particles, and we find the $b\bar{b}$ and $2h_1$ constraints are always satisfied.  In FIG.~\ref{DM:Detect}(b) we show the $W^+W^-$ constraint in the $m_\chi$-$v_s$ plane for the two scalar mass sets in Eq.~(\ref{eq:benchmark_masses}).  As shown in the plot, the indirect detection constraints are mostly caused by heavy scalar resonances near $m_\chi=140,210,300$~GeV (black dashed lines).

Before concluding this section, we summarize the DM relic, direct, and indirect detection constraints in FIG.~\ref{DM:all}.

\begin{figure}[t]
	\includegraphics[width=0.4\textwidth]{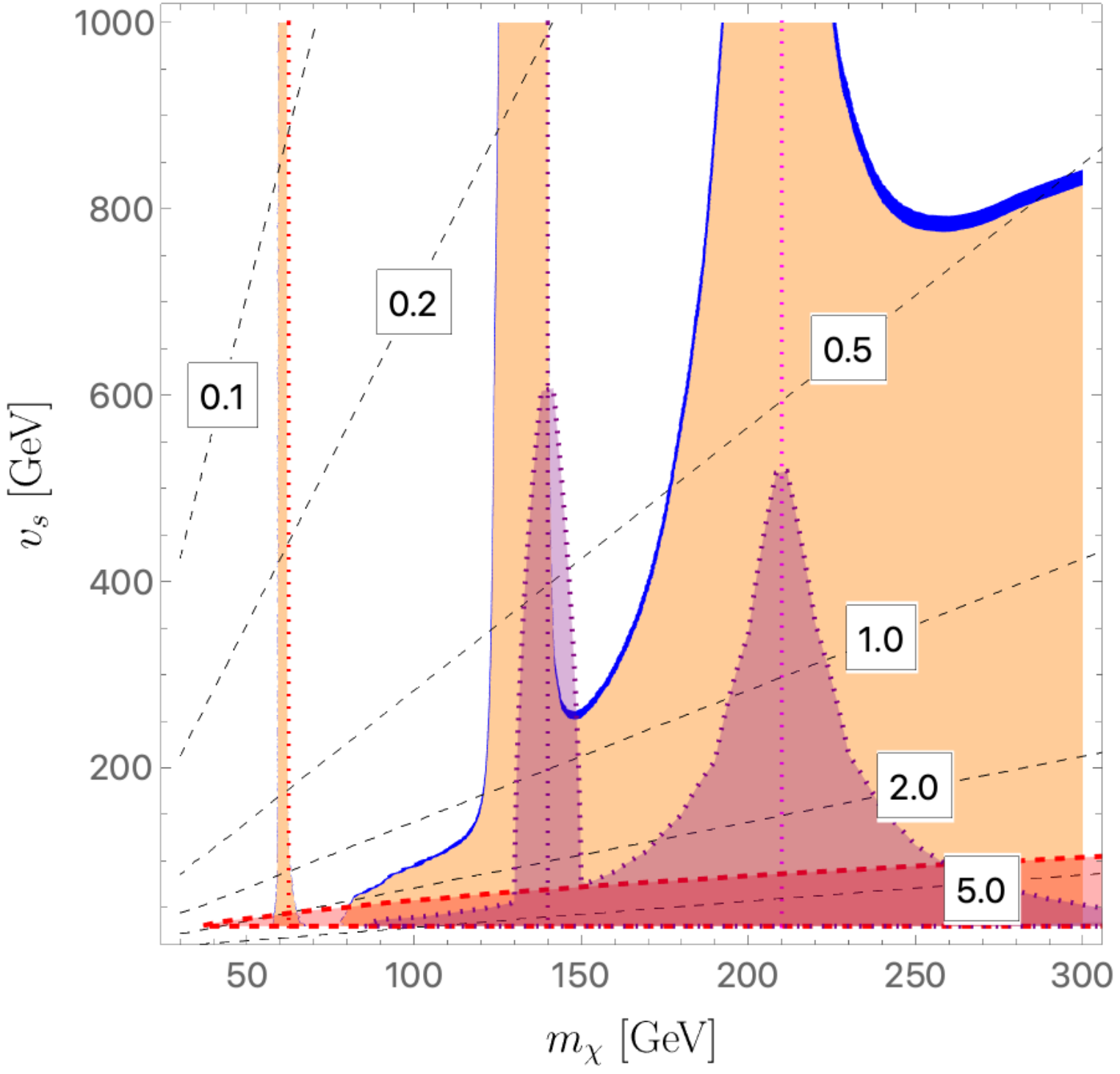}
	\includegraphics[width=0.4\textwidth]{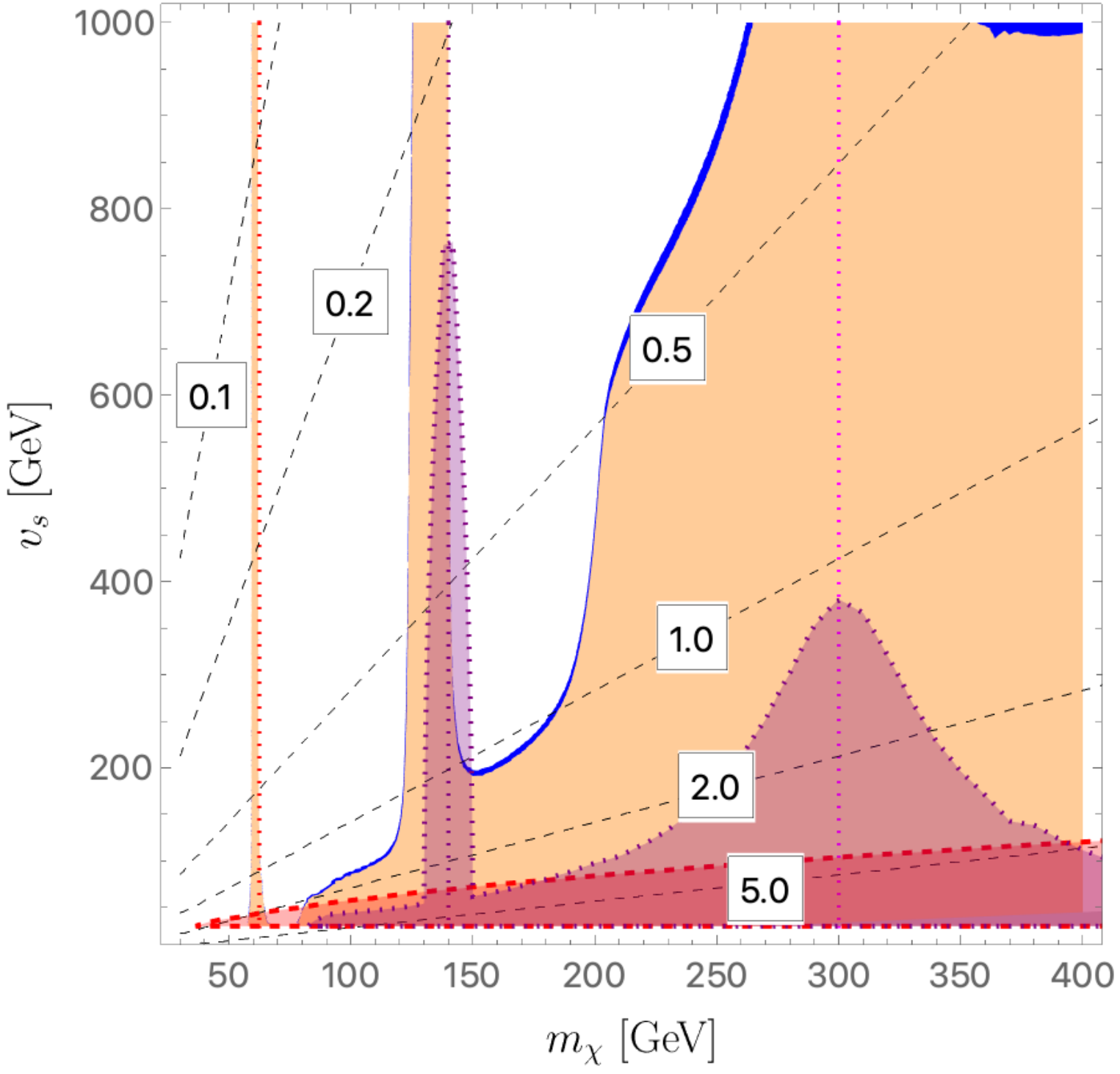}
\\
\vspace{-0.4cm}
(a) \hspace{6cm} (b)
\caption{\label{DM:all} {Summary of the DM relic, direct, and indirect detections in the $m_\chi$-$v_s$ plane for scalar mass sets (a) 1 and (b) 2. The orange and blue regions, black dashed contours, and vertical red, purple, and magenta lines are of the same meanings as in FIG.~\ref{DMrelic}. We shade the region excluded by the direct detection in red with dashed boundaries, and the region excluded by the indirect detection in purple with dotted boundaries.}}
\end{figure}

\section{Triple and Quadruple Higgs Productions at the LHC}\label{sec:4}

After considering  current experimental constraints on our model, in this section we propose four benchmarks and consider their collider phenomenology at the 14-TeV LHC. To reduce the number of free parameters, we turn off the cubic couplings for $S$, $c_1$ and $c_2$ in Eq.~(\ref{L:Scalar}). We choose to focus on the possibility that $h_2$ is mostly CP-even and $h_3$ mostly CP-odd, which can be achieved by setting $\theta_{12}+\theta_{23}=0$. With this parameter choice, ${g}_{223}$ is completely determined and we are free to set $b_2=d_3=0$. In this scenario, CPV takes place in the Higgs-to-Higgs decays in the $3h_1$ final state through the ${g}_{123}$ coupling and in the $4h_1$ final state through the ${g}_{223}$ coupling, as can be seen from Eq.~(\ref{eq:gijkcoupling}).\footnote{We could consider the other scenario where $h_3$ is mostly CP-even and $h_2$ is mostly CP-odd. In this case   the terms proportional to $s_{12+23}$ in ${g}_{223}$ could be cancelled by properly choosing $b_2$ and $d_3$, leading  to similar $3h_1/4h_1$ decay characteristics.} Therefore, the triple and quadruple Higgs productions at the LHC could be smoking gun signatures of CPV in the model.

In Table~\ref{BP:param} we propose the four benchmark scenarios, \{BP1, BP2, BP3, BP4\}, to further study the $3h_1/4h_1$ signatures at the LHC. These four benchmarks have different collider phenomenology: BP1 and BP3 are chosen to allow the the $3h_1$ production, while BP2 and BP4 are chosen to afford both the $3h_1$ and $4h_1$ productions. They satisfy all experimental constraints considered in previous sections. We fix all the parameters except $\delta_2$, so as to look for regions of parameter space which maximize the event rates of $3h_1/4h_1$ final states. In this regard, we need to suppress the $h_2,h_3\to\chi\overline{\chi}$ decays by choosing the appropriate $m_\chi$ and $v_s$, resulting in interesting interplay with the DM relic density which we explain as follows.

\begin{table}[t]
\centering
\begin{tabular}{>{\centering\arraybackslash}p{4cm}|>{\centering\arraybackslash}p{4cm}|>{\centering\arraybackslash}p{4cm}|>{\centering\arraybackslash}p{4cm}}
\toprule
BP1 & BP3 & BP2 & BP4 \\
\toprule
\multicolumn{4}{>{\centering\arraybackslash}p{16.0cm}}{$m_{h_2}=280$~GeV, $\epsilon=0.1$, $b_2=c_1=c_2=d_3=0$, $\theta_{\delta_3}=-\frac{\pi}{2}+2(\theta_{12}+\theta_{23})$, $\vert\delta_3\vert=3.5$} \\
\colrule
\multicolumn{2}{>{\centering\arraybackslash}p{8.0cm}|}{$m_{h_3}=420$~GeV, ~$\theta_{12}=0.73$, ~$\theta_{23}=-0.73$} & 
\multicolumn{2}{>{\centering\arraybackslash}p{8.0cm}}{$m_{h_3}=600$~GeV, ~$\theta_{12}=0.41$, ~$\theta_{23}=-0.41$} \\
\colrule
$m_\chi=280$~GeV, ~$v_s=200$~GeV & $m_\chi=187$~GeV, ~$v_s=241$~GeV & $m_\chi=420$~GeV, ~$v_s=200$~GeV & $m_\chi=156$~GeV, ~$v_s=200$~GeV \\
\toprule
\multicolumn{4}{>{\centering\arraybackslash}p{16.0cm}}{Free Parameter: $\delta_2$} \\
\botrule
\end{tabular}
\caption{\label{BP:param} The parameters of the four BPs; $\delta_2$ is the only remaining free parameter.}
\end{table}

In BP1 and BP2, we choose somewhat heavier DM masses, $(m_\chi, v_s)=(280, 200)$~GeV for BP1 and (420, 200)~GeV for BP2, both of which are heavier than $m_{h_3}/2$ in their respective benchmarks so that  $h_2, h_3\to \bar{\chi}\chi$ decays are forbidden and the Higgs-to-Higgs decay branching fractions are maximized. However, in this case $\chi$ pairs   annihilate efficiently into $h_1h_2$ (94\%) and $2h_2$ (4\%), resulting in a vanishingly small relic density:
\begin{equation}
	\Omega_{\chi}^{\rm BP1, BP2} h^2\sim10^{-4} ~.
\end{equation}
Other DM candidates (such as the axions) need to be present in these two benchmarks to satisfy the relic density. It is possible to choose benchmarks which fully account for the DM relic density with lighter DM masses, and they are presented in BP3 and BP4, where we choose $(m_\chi, v_s)=(187, 241)$~GeV for BP3 and (156, 200)~GeV for BP4.
In BP3, $\chi \overline\chi$ pairs mainly annihilate to $WW$ (61\%), $ZZ$ (28\%), and $t\overline{t}$ (8\%), while in BP4, they mainly annihilate to $WW$ (61\%), $ZZ$ (27\%), and $2h_1$ (12\%), giving the DM relic densities
\begin{equation}
	\Omega_{\chi}^{\rm BP3} h^2=0.121 ~,~ \Omega_{\chi}^{\rm BP4} h^2=0.124 ~.
\end{equation}

\begin{figure}[t]
\centering
	\includegraphics[width=0.45\textwidth]{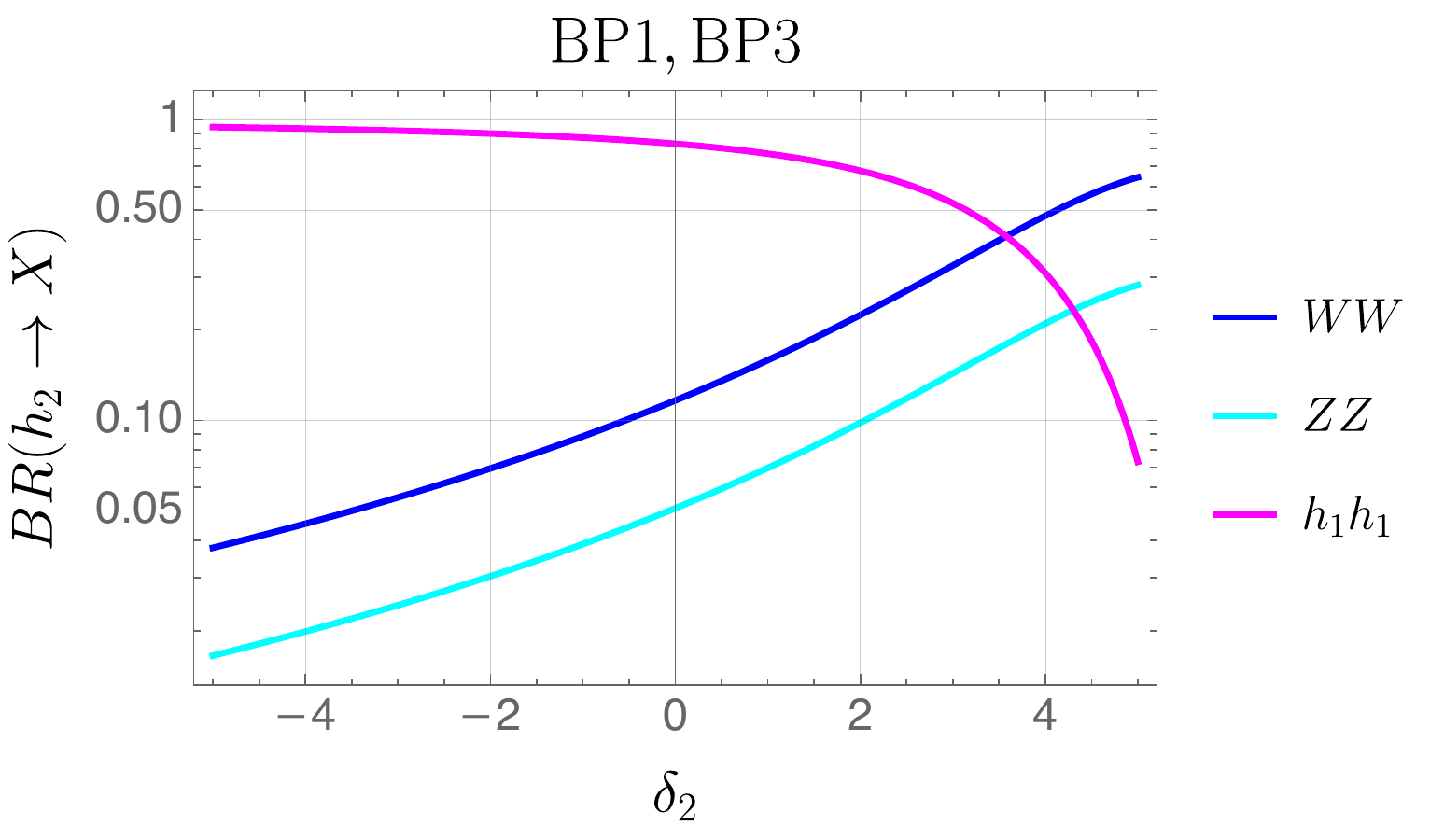}
	\includegraphics[width=0.45\textwidth]{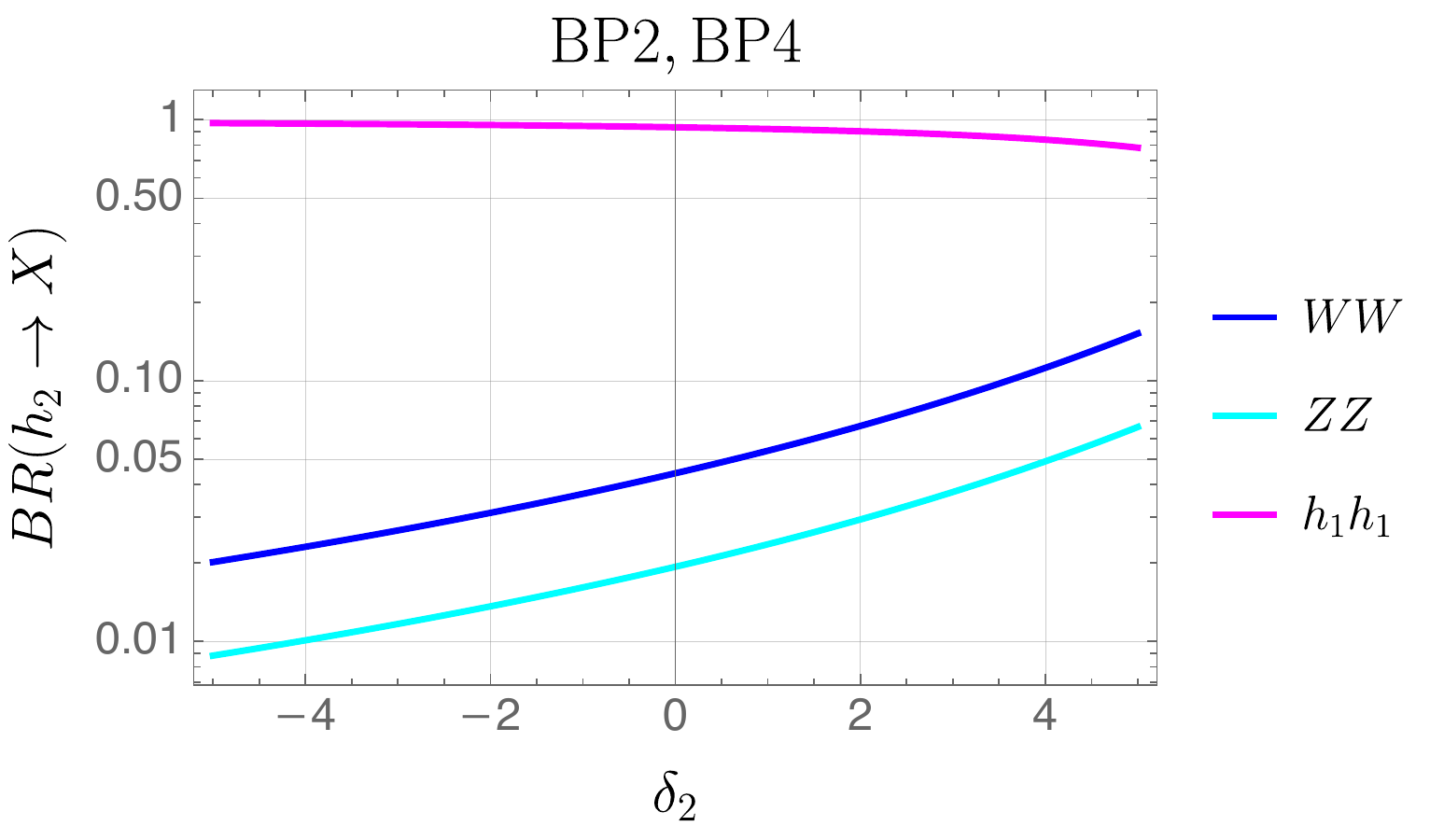}
\\
\vspace{-0.4cm}
(a) \hspace{6.8cm} (b)
\\
\vspace{0.3cm}
	\includegraphics[width=0.45\textwidth]{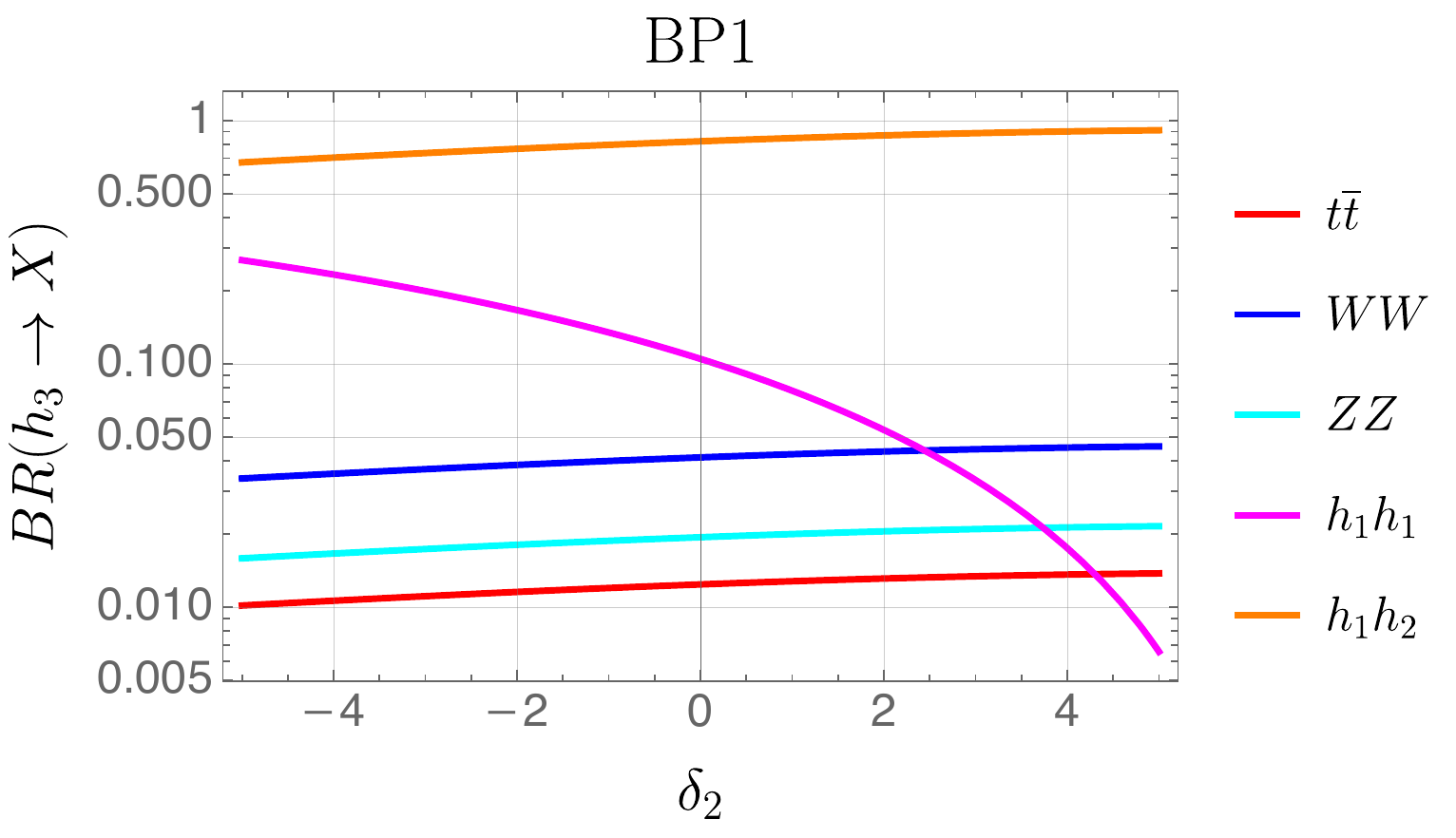}
	\includegraphics[width=0.45\textwidth]{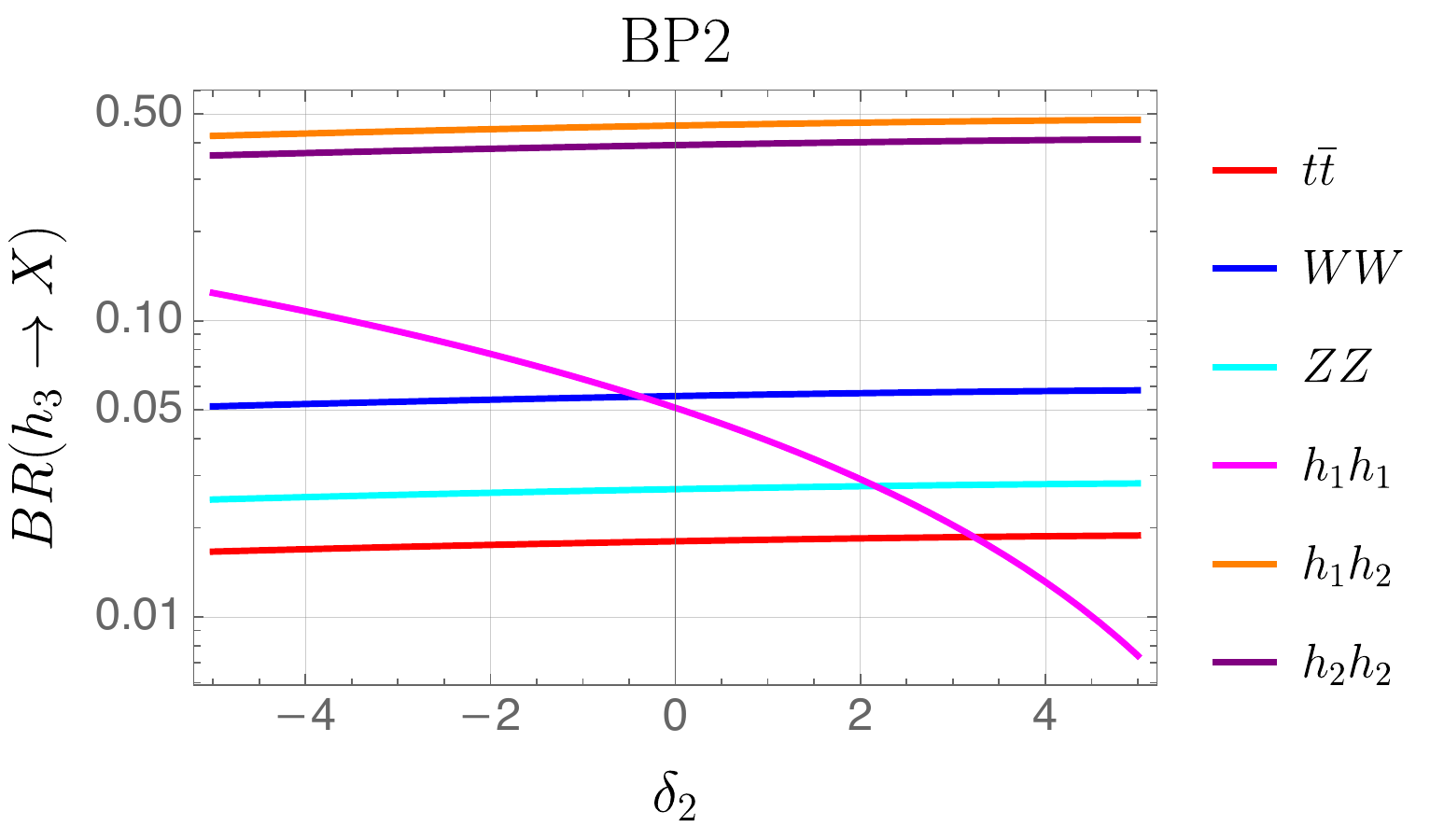}
\\
\vspace{-0.4cm}
(c) \hspace{6.8cm} (d)
\\
\vspace{0.3cm}
	\includegraphics[width=0.45\textwidth]{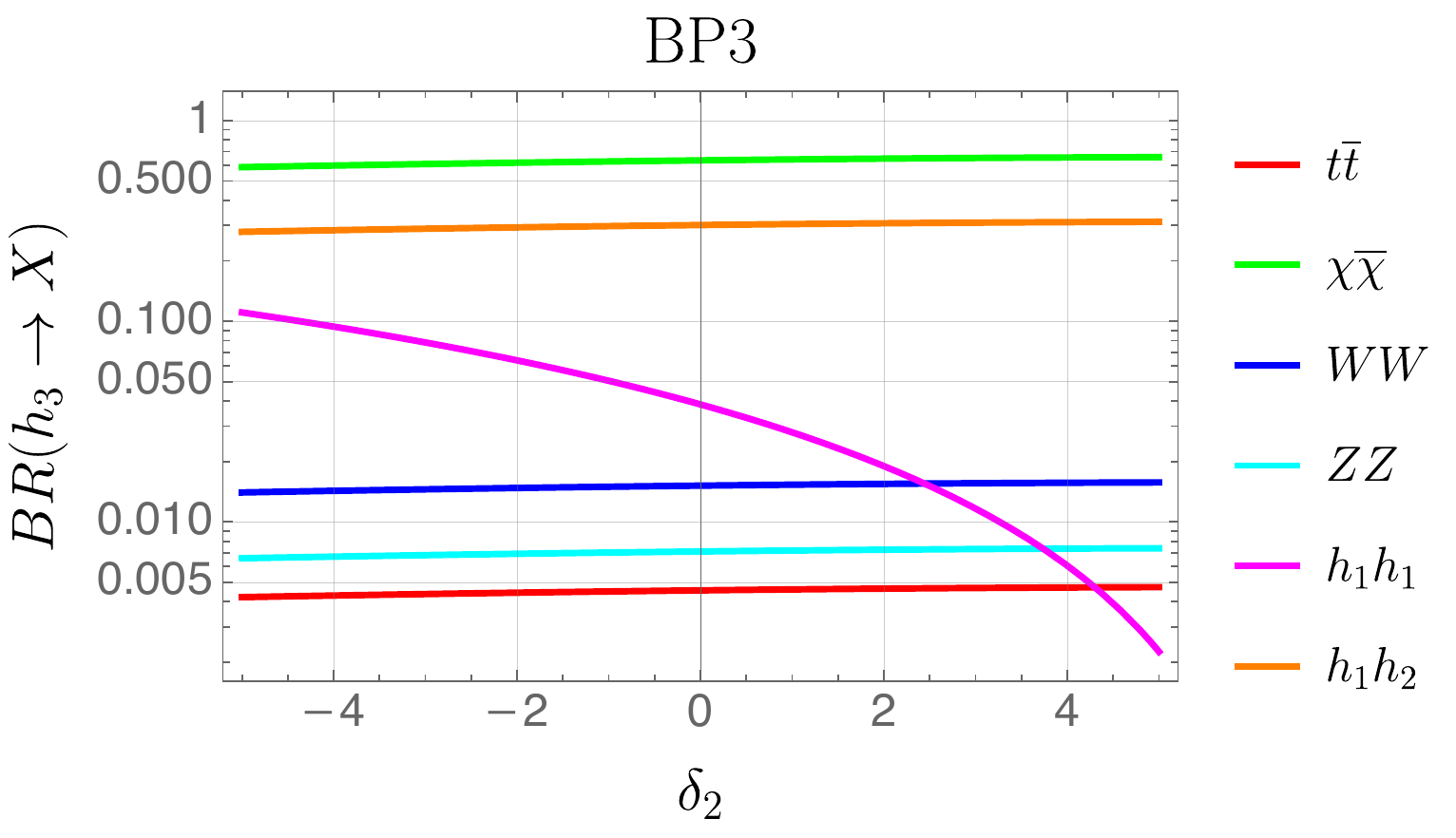}
	\includegraphics[width=0.45\textwidth]{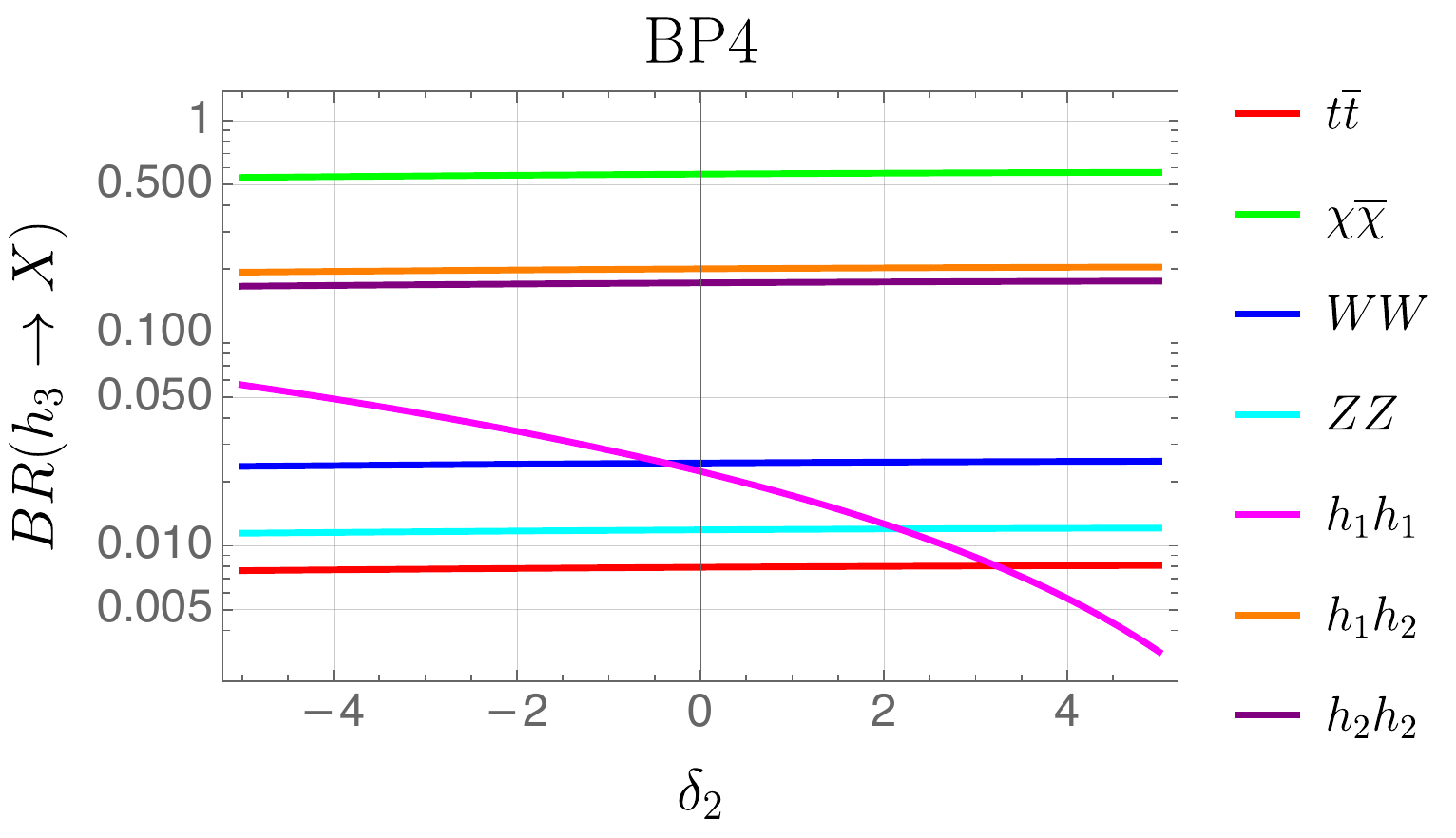}
\\
\vspace{-0.4cm}
(e) \hspace{6.8cm} (f)
\caption{\label{BRs:h2h3:34} Decay BRs of $h_2$ for (a) BP1/BP3 and for (b) BP2/BP4, and of $h_3$ for (c) BP1, for (d) BP2, for (e) BP3, and for (f) BP4, as functions of $\delta_2$.}
\end{figure}

The production of the heavy scalars at the 14-TeV LHC goes through the ggF channel and the cross-sections are 
\begin{align}
\begin{split}
\sigma_{\rm ggF}(pp\to h_2/h_3)_{\rm BP1,BP3} =55~{\rm fb} ~,
\\ 
\sigma_{\rm ggF}(pp\to h_2/h_3)_{\rm BP2,BP4} =20~{\rm fb} ~,
\end{split}
\end{align}
where we have chosen the $\theta_{12}$ values in such a way that the cross sections for $h_2$ and $h_3$ are the same, as mentioned in Sec.~\ref{sec:3:3}.
The decay branching ratios (BRs) of the heavy scalars are plotted against $\delta_2$, the only free parameter in our benchmarks, in FIG.~\ref{BRs:h2h3:34}. 
We note that the decay BRs of $h_2$ in BP3 are the same as those in BP1, and those in BP4 the same as those in BP2.  This is because the $h_2\to\chi\overline{\chi}$ decay remains forbidden in BP3 and BP4 and the partial widths of other decay channels remain unchanged.  Furthermore, as long as the scalar mixing angles are fixed, the decay partial widths of the $h_i\to \bar{f}_{\rm SM} f_{\rm SM}$ are also fixed in the benchmarks.

Given the production cross-sections and the decay BRs, we show in FIG.~\ref{sigmas:34} the event rate of $3h_1/4h_1$ final states as a function of $\delta_2$.
{In BP1/BP3, the maximum event rates for the $3h_1$  are obtained for $\delta_2=-1.13$ and $-3.27$}, respectively, while in BP2/BP4, the maximum rates for $3h_1$ and $4h_1$ final states take place under different conditions. Maximizing the $4h_1$ rate leads to $\delta_2=-1.92$ in BP2 and $\delta_2=-4.56$ in BP4. To summarize,
\begin{equation}
\begin{gathered}
	\sigma_{\rm ggF}(pp\to h_3\to3h_1)^{\rm BP1}_{\rm max}=38.2~{\rm fb} ~, \\
	\sigma_{\rm ggF}(pp\to h_3\to3h_1)^{\rm BP3}_{\rm max}=14.6~{\rm fb} ~, \\
	\sigma_{\rm ggF}(pp\to h_3\to3h_1)^{\rm BP2}_{\rm max}=8.60~{\rm fb} ~,~
	\sigma_{\rm ggF}(pp\to h_3\to4h_1)^{\rm BP2}_{\rm max}=7.05~{\rm fb} ~, \\
	\sigma_{\rm ggF}(pp\to h_3\to3h_1)^{\rm BP4}_{\rm max}=3.82~{\rm fb} ~,~
	\sigma_{\rm ggF}(pp\to h_3\to4h_1)^{\rm BP4}_{\rm max}=3.18~{\rm fb} ~.
\end{gathered}
\end{equation}

\begin{figure}[t]
\centering
	\includegraphics[width=0.45\textwidth]{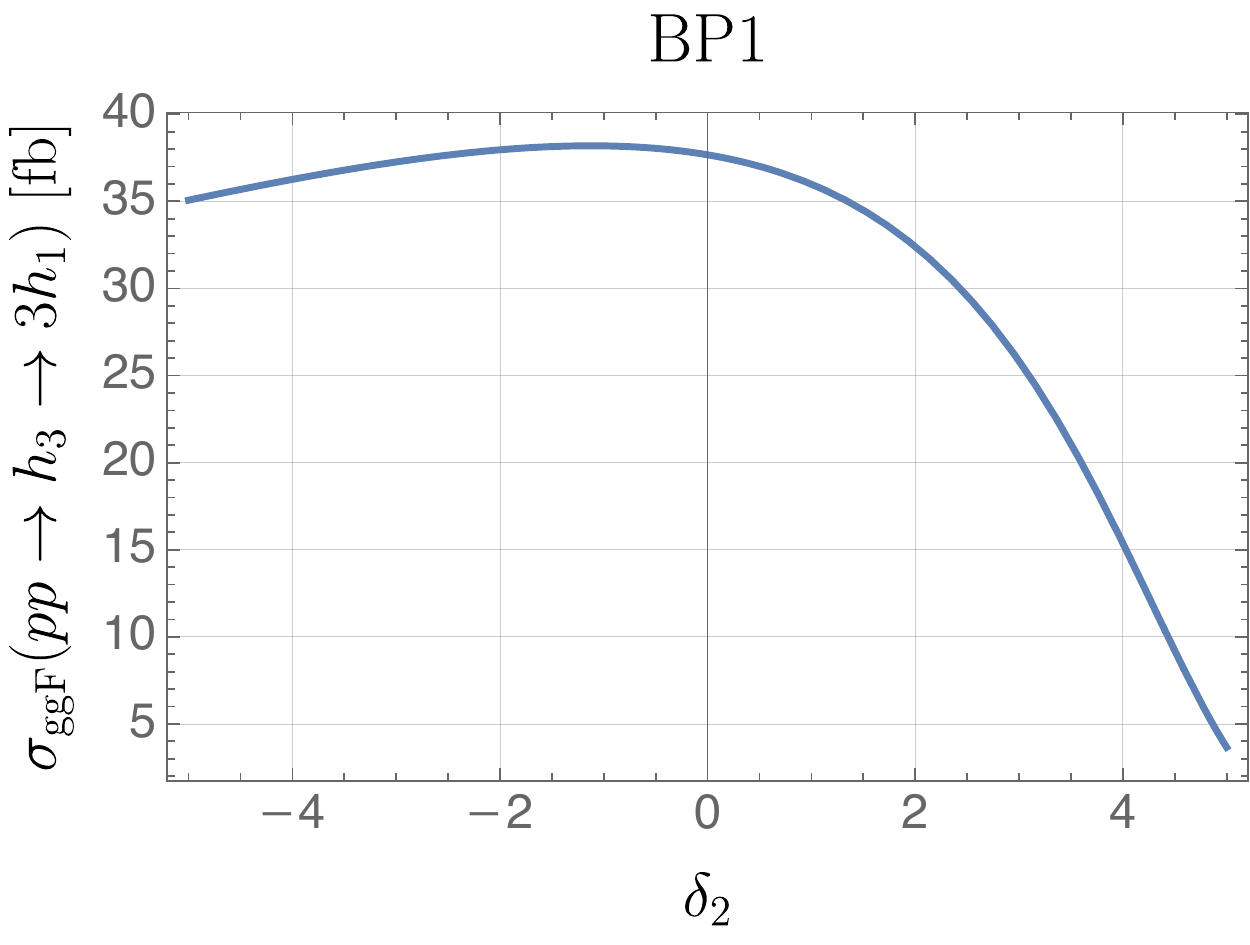}
	\includegraphics[width=0.45\textwidth]{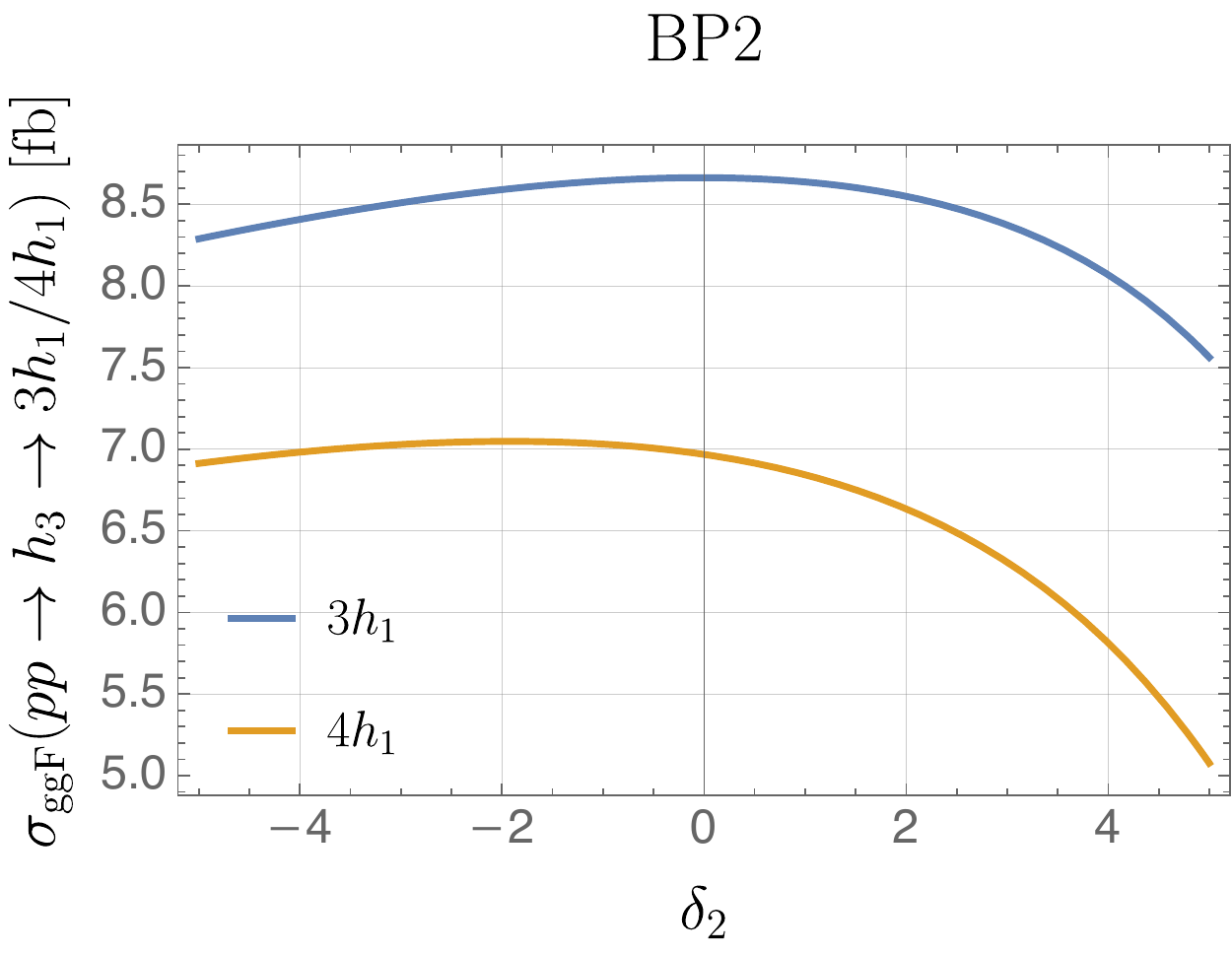}
\\
\vspace{-0.4cm}
(a) \hspace{6.8cm} (b)
\\
\vspace{0.3cm}
	\includegraphics[width=0.45\textwidth]{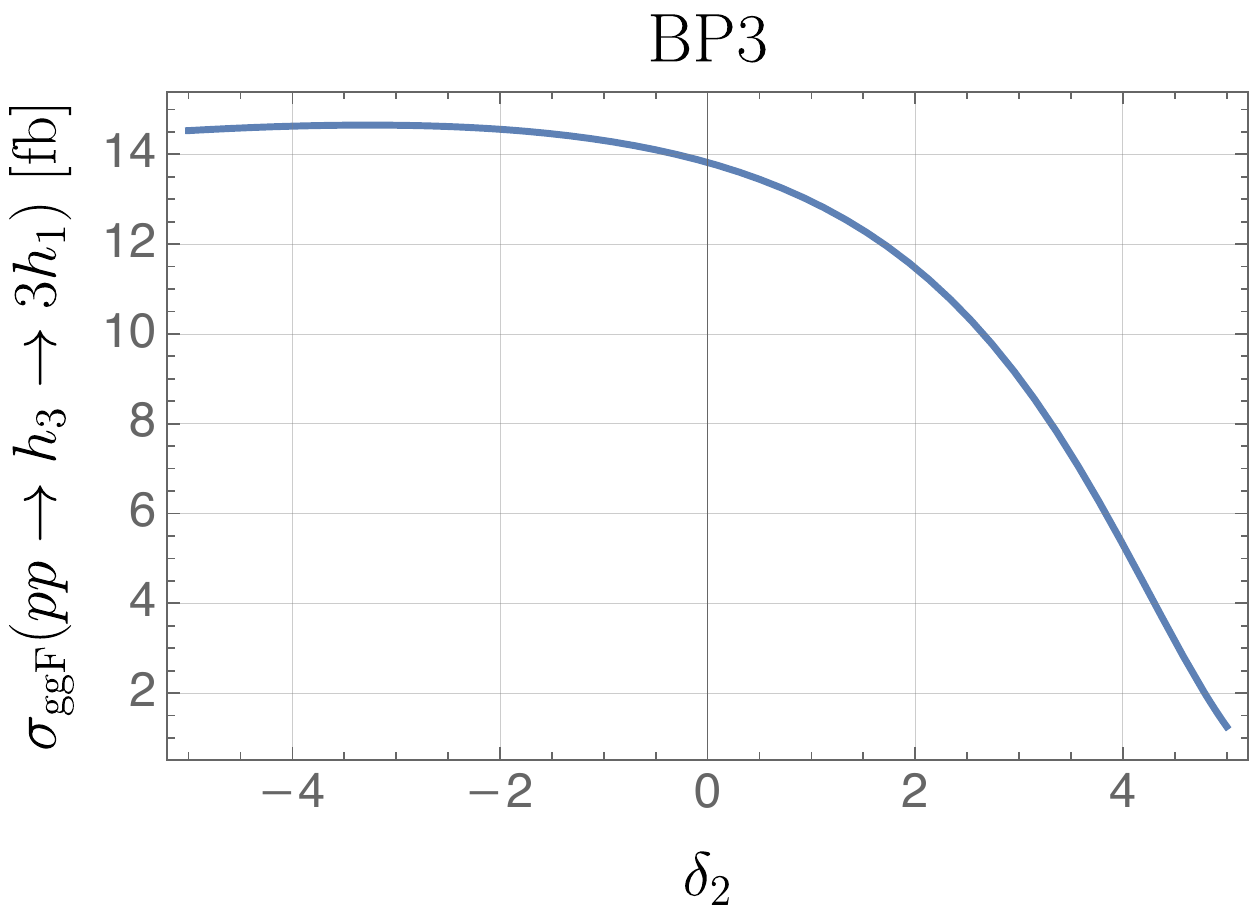}
	\includegraphics[width=0.45\textwidth]{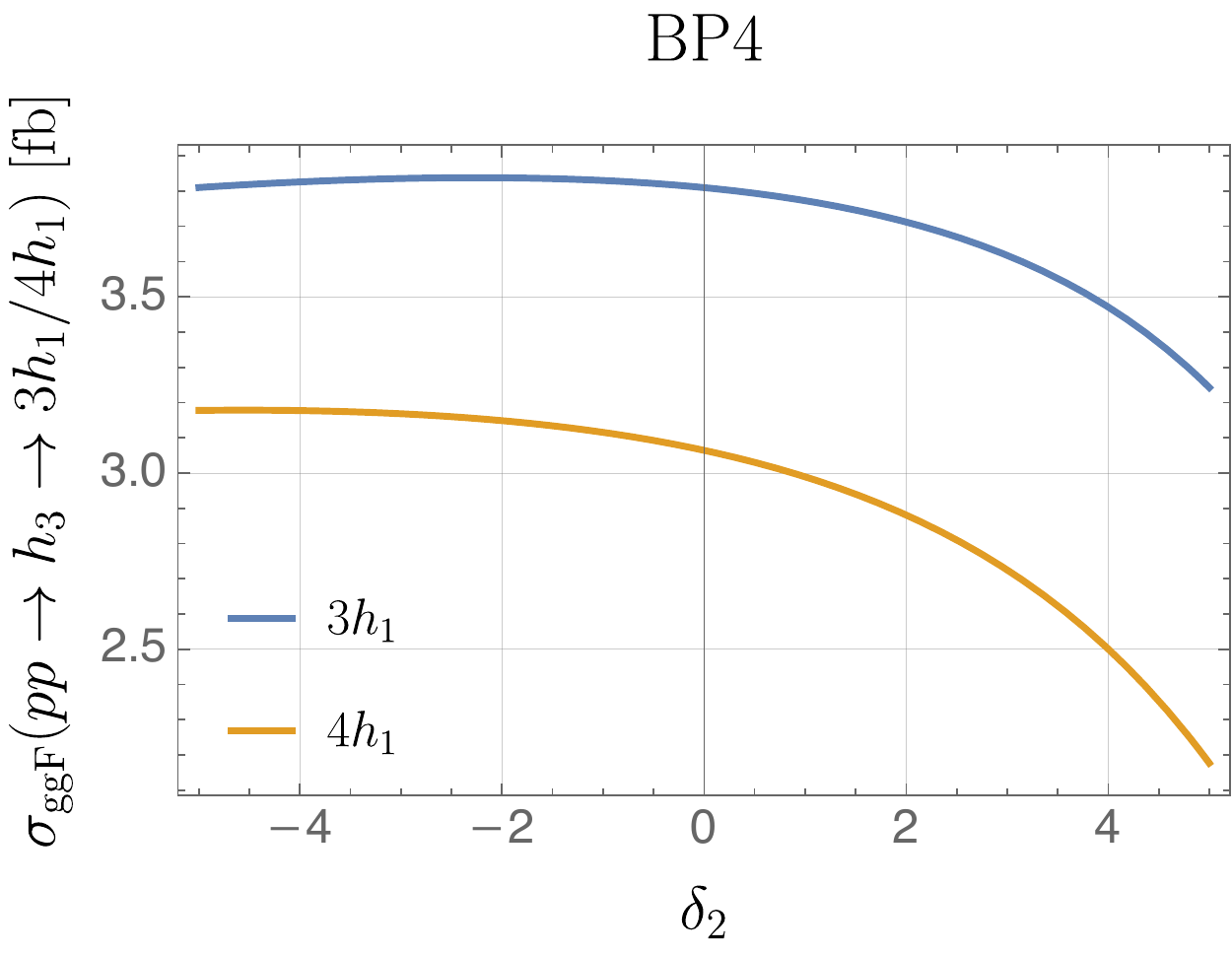}
\\
\vspace{-0.4cm}
(c) \hspace{6.8cm} (d)
\caption{\label{sigmas:34} $\sigma_{\rm ggF}(pp\to h_3\to 3h_1/4h_1)$ versus $\delta_2$ for (a) BP1, (b) BP2, (c) BP3, and (d) BP4.}
\end{figure}

With event rates for $3h_1$ and $4h_1$ around $\mathcal{O}(10)$ and $\mathcal{O}(1)$~fb, respectively, it is imperative to dedicate experimental efforts to search for these final states at the 14-TeV LHC. 

\section{Conclusions}\label{sec:5}

In this work we have proposed a simple model of CP violation and dark matter, where the dark matter is a vector-like ``dark fermion'' $(\bar{\chi},\chi)$ which interacts with the SM only through a messenger scalar $S$ that is an electroweak singlet. New sources of CPV arise in the most general potential for the Higgs doublet $H$ and the singlet $S$ as well as the dark Yukawa coupling between $S$ and the dark matter.  We have shown that such a simple setup could satisfy all current experimental constraints: Higgs signal strength measurements and searches for new neutral scalars at the LHC, precision electroweak measurements, electron EDM constraints, DM relic density, and DM direct and indirect detections.  Notably, there is no new contributions to the electron EDM up to the two-loop level due to the fact that there is no new sources of CPV entering the visible fermion sector.

Novel signatures of CPV in this model come from Higgs-to-Higgs decays, $h_3\to h_1h_2$, which involves a CPV coupling and does not vanish even in the exact alignment limit when the 125-GeV Higgs is exactly SM-like, in sharp contrast to the C2HDMs where the corresponding coupling becomes zero in the alignment limit.  Moreover, the Higgs-to-Higgs decays, which include $h_3\to 2h_2$ and $h_2\to 2h_1$, can give rise to yet-to-be-searched-for final states such as triple and quadruple 125-GeV Higgs bosons, which are highly suppressed within the SM. There is also no anomalous couplings in the Higgs sector, since $S$ is a singlet and does not couple to SM fermions.  Only the 125-GeV Higgs coupling strengths are reduced due to the mass mixing.

While the $3h_1$ and $4h_1$ final states are smoking-gun signatures of CPV in the model (and in C2HDMs as well), it is conceivable that more complicated extensions of the SM without CPV, such as 2HDMs with an additional real singlet scalar, could also give rise to similar final states. In this regard, we point out that these more complicated extensions require the presence of additional neutral or charged scalars that are not present in the CPV models, which could be used to distinguish the models. Furthermore, it may be possible to unambiguously detect the presence of CPV trilinear scalar couplings through interference effects in the three-body decay as described in Ref.~\cite{Chen:2014ona}, by considering the SM production of $3h_1$ interfering with $h_3\to 3h_1$ in the off-shell region, which is beyond the scope of the present work. It would also be interesting to consider ways to detect the CPV in the dark Yukawa coupling via, for example, directional direct detection.

Whether this model can accommodate the observed baryon asymmetry in the Universe remains to be seen. The feature that new sources of CPV are associated with interactions of the messenger scalar with the dark matter and the Higgs boson may indicate a connection between the proximity between the observed baryon relic abundance and the dark matter relic abundance~\cite{Kaplan:1991ah,Kitano:2004sv,Kaplan:2009ag}. 

We hope it is clear that our work opens up a new frontier of searching for multi-Higgs bosons at the LHC. A detailed study on the discovery potential of the $3h_1/4h_1$ final states at the LHC is obviously necessary, which we plan to pursue in the future.

\section*{Acknowledgments}

We thank Marcela Carena, Jia Liu, Carlos Wagner and Xiaoping Wang for their comments on the manuscript.  We also acknowledge helpful discussions with Xiaoping Wang on the EDM constraint issues. IL would like to thank the support and the hospitality of the Physics Division of National Center for Theoretical Sciences (NCTS), Taiwan, where this project was initiated.  TKC was supported in part by the grant of NCTS.  CWC was supported in part by the Ministry of Science and Technology, Taiwan under the Grant No.~MOST-108-2112-M-002-005-MY3.  IL is supported in part by the U.S. Department of Energy under contracts No. DE- AC02-06CH11357 at Argonne and No. DE-SC0010143 at Northwestern.

\appendix

\section{List of Couplings}\label{sec:a}

We list in this appendix the trilinear and quartic scalar couplings as well as the couplings of the scalar fields to the SM fermions and weak gauge bosons in the model:

\begin{itemize}
\item
Trilinear Couplings:
\begin{align}
	{g}_{111} &= -3\frac{m_{h_1}^2}{v}+\mathcal{O}(\epsilon^2) ~, \\
	{g}_{122} &= -\frac{v}{2}\Big[ \delta_2+{\rm Re}\big(\delta_3e^{-2i(\theta_{12}+\theta_{23})}\big) \Big] + \mathcal{O}(\epsilon) ~, \\
	{g}_{133} &= -\frac{v}{2}\Big[ \delta_2-{\rm Re}\big(\delta_3e^{-2i(\theta_{12}+\theta_{23})}\big) \Big] + \mathcal{O}(\epsilon) ~, \\
	{g}_{123} &= -\frac{v}{2}{\rm Im}\big(\delta_3e^{-2i(\theta_{12}+\theta_{23})}\big) + \mathcal{O}(\epsilon) ~, \\
	{g}_{223} &= -\frac{1}{12v_s}\Big\{ 3s_{12+23}\big[ -8m_{h_2}^2 + 4m_{h_3}^2 + v^2\delta_2 - 4b_2(1+3c_{2(12+23)}) + 3v_s^2s_{2(12+23)}{\rm Im} d_3 \big] \nonumber \\
	&\quad - 9v^2{\rm Im}\big(\delta_3e^{-3i(\theta_{12}+\theta_{23})}\big)-\sqrt{2}v_s{\rm Im}\big(3c_1e^{-3i(\theta_{12}+\theta_{23})} - 2c_2e^{i(\theta_{12}+\theta_{23})} \nonumber \\
	&\quad +3c_2e^{3i(\theta_{12}+\theta_{23})}\big) \Big\} + \mathcal{O}(\epsilon) ~, \\
	{g}_{233} &= -\frac{1}{12v_s}\Big\{ 3c_{12+23}\big[ 8m_{h_3}^2 - 4m_{h_2}^2 - v^2\delta_2 + 4b_2(1-3c_{2(12+23)}) + 3v_s^2s_{2(12+23)}{\rm Im} d_3 \big] \nonumber \\
	&\quad + 9v^2{\rm Re}\big(\delta_3e^{-i3(\theta_{12}+\theta_{23})}\big)+\sqrt{2}v_s{\rm Re}\big(3c_1e^{-3i(\theta_{12}+\theta_{23})} - 2c_2e^{i(\theta_{12}+\theta_{23})} \nonumber \\
	&\quad -3c_2e^{3i(\theta_{12}+\theta_{23})}\big) \Big\} + \mathcal{O}(\epsilon) ~, \\
	{g}_{222} &= \frac{1}{4v_s}\Big[ 3c_{12+23}(-4m_{h_2}^2+v^2\delta_2+8b_2s_{12+23}^2) + 3v^2{\rm Re}\big(\delta_3e^{-3i(\theta_{12}+\theta_{23})}\big) \nonumber \\
	&\quad +\sqrt{2}v_s{\rm Re}\big(c_1e^{-3i(\theta_{12}+\theta_{23})} - c_2e^{3i(\theta_{12}+\theta_{23})} + 2c_2e^{i(\theta_{12}+\theta_{23})}\big) \nonumber \\
	&\quad- 6v_s^2s_{12+23}^3{\rm Im} d_3 \Big] + \mathcal{O}(\epsilon) ~, \\
	{g}_{333} &= -\frac{1}{4v_s}\Big[ 3s_{12+23}(-4m_{h_3}^2+v^2\delta_2+8b_2c_{12+23}^2) + 3v^2{\rm Im}\big(\delta_3e^{-3i(\theta_{12}+\theta_{23})}\big) \nonumber \\
	&\quad +\sqrt{2}v_s{\rm Im}\big(c_1e^{-3i(\theta_{12}+\theta_{23})} + c_2e^{3i(\theta_{12}+\theta_{23})} + 2c_2e^{i(\theta_{12}+\theta_{23})}\big) \nonumber \\
	&\quad + 6v_s^2c_{12+23}^3{\rm Im} d_3 \Big] + \mathcal{O}(\epsilon) ~, \\
	{g}_{112} &= \frac{\epsilon}{v}\Big\{\big(2m_{h_1}^2+m_{h_2}^2\big)s_{12}-v^2\big[\delta_2s_{12}+{\rm Im}\big(\delta_3e^{-i(\theta_{12}+2\theta_{23})}\big)\big]\Big\} + \mathcal{O}(\epsilon^2) ~, \\
	{g}_{113} &= \frac{\epsilon}{v}\Big\{\big(2m_{h_1}^2+m_{h_3}^2\big)c_{12}-v^2\big[\delta_2c_{12}-{\rm Re}\big(\delta_3e^{-i(\theta_{12}+2\theta_{23})}\big)\big]\Big\} + \mathcal{O}(\epsilon^2) ~,
\end{align}

\item
Quartic Couplings:
\begin{align}
	{g}_{1111} &= -3\frac{m_{h_1}^2}{v^2} + \mathcal{O}(\epsilon^2) ~, \\
	{g}_{1122} &= -\frac{1}{2}\Big[ \delta_2+{\rm Re}\big(\delta_3e^{-2i(\theta_{12}+\theta_{23})}\big) \Big] + \mathcal{O}(\epsilon^2) ~, \\
	{g}_{1133} &= -\frac{1}{2}\Big[ \delta_2-{\rm Re}\big(\delta_3e^{-2i(\theta_{12}+\theta_{23})}\big) \Big] + \mathcal{O}(\epsilon^2) ~, \\
	{g}_{1123} &= -\frac{1}{2}{\rm Im}\big(\delta_3e^{-2i(\theta_{12}+\theta_{23})}\big) + \mathcal{O}(\epsilon^2) ~, \\
	{g}_{2233} &= \frac{1}{48v_s^2}\Big\{ -24(m_{h_2}^2+m_{h_3}^2)+72(m_{h_2}^2-m_{h_3}^2)c_{2(12+23)}+24b_2(1+3c_{4(12+23)}) \nonumber \\
	&\quad +12v^2\big[ \delta_2-3{\rm Re}\big(\delta_3e^{-4i(\theta_{12}+\theta_{23})}\big) \big] - 4\sqrt{2}v_s{\rm Re}\big( 9c_1e^{-4i(\theta_{12}+\theta_{23})}-4c_2 \nonumber \\
	&\quad -3c_2e^{4i(\theta_{12}+\theta_{23})} \big) - 9v_s^2{\rm Re}\big( 3d_3e^{-4i(\theta_{12}+\theta_{23})} - d_3e^{4i(\theta_{12}+\theta_{23})} - 2d_3  \big) \Big\} + \mathcal{O}(\epsilon) ~, \\
	{g}_{2223} &= -\frac{1}{16v_s^2}\Big[ -24(m_{h_2}^2-m_{h_3}^2)s_{2(12+23)}-24b_2s_{4(12+23)} - 12v^2{\rm Im}\big(\delta_3e^{-4i(\theta_{12}+\theta_{23})}\big) \nonumber \\
	&\quad -4\sqrt{2}v_s{\rm Im}\big(3c_1e^{-4i(\theta_{12}+\theta_{23})}+c_2e^{4i(\theta_{12}+\theta_{23})}\big) + 3v_s^2{\rm Im}\big(4d_3e^{-2i(\theta_{12}+\theta_{23})}\nonumber \\
	&\quad -3d_3e^{-4i(\theta_{12}+\theta_{23})}-d_3e^{4i(\theta_{12}+\theta_{23})}\big) \Big] + \mathcal{O}(\epsilon) ~, \\
	{g}_{2333} &= \frac{1}{16v_s^2}\Big[ -24(m_{h_2}^2-m_{h_3}^2)s_{2(12+23)}-24b_2s_{4(12+23)} - 12v^2{\rm Im}\big(\delta_3e^{-4i(\theta_{12}+\theta_{23})}\big) \nonumber \\
	&\quad -4\sqrt{2}v_s{\rm Im}\big(3c_1e^{-4i(\theta_{12}+\theta_{23})}+c_2e^{4i(\theta_{12}+\theta_{23})}\big) - 3v_s^2{\rm Im} \big(4d_3e^{-2i(\theta_{12}+\theta_{23})}\nonumber \\
	&\quad +3d_3e^{-4i(\theta_{12}+\theta_{23})}+d_3e^{4i(\theta_{12}+\theta_{23})}\big) \Big] + \mathcal{O}(\epsilon) ~, \\
	{g}_{2222} &= \frac{1}{4v_s^2}\Big\{ -6(m_{h_2}^2+m_{h_3}^2)-6(m_{h_2}^2-m_{h_3}^2)c_{2(12+23)}+3v^2\big[\delta_2+{\rm Re}\big(\delta_3e^{-4i(\theta_{12}+\theta_{23})}\big)\big] \nonumber \\
	&\quad +12b_2s_{2(12+23)}^2 + \sqrt{2}v_s\big[{\rm Re}\big(3c_1e^{-4i(\theta_{12}+\theta_{23})}+4c_2-c_2e^{4i(\theta_{12}+\theta_{23})}\big) \big] \nonumber \\
	&\quad -6v_s^2s_{12+23}^3{\rm Im}\big(3d_3e^{-i(\theta_{12}+\theta_{23})}+d_3e^{i(\theta_{12}+\theta_{23})}\big) \Big\} + \mathcal{O}(\epsilon) ~, \\
	{g}_{3333} &= \frac{1}{4v_s^2}\Big\{ -6(m_{h_2}^2+m_{h_3}^2)-6(m_{h_2}^2-m_{h_3}^2)c_{2(12+23)}+3v^2\big[\delta_2+{\rm Re}\big(\delta_3e^{-4i(\theta_{12}+\theta_{23})}\big)\big] \nonumber \\
	&\quad +12b_2s_{2(12+23)}^2 + \sqrt{2}v_s\big[{\rm Re}\big(3c_1e^{-4i(\theta_{12}+\theta_{23})}+4c_2-c_2e^{4i(\theta_{12}+\theta_{23})}\big) \big] \nonumber \\
	&\quad +6v_s^2c_{12+23}^3{\rm Re}\big( 3d_3e^{-i(\theta_{12}+\theta_{23})}-d_3e^{i(\theta_{12}+\theta_{23})} \big) \Big\} + \mathcal{O}(\epsilon) ~, \\
	{g}_{1112} &= \frac{3}{2}\epsilon\Bigg[\Bigg(\frac{2m_{h_1}^2}{v^2}-\delta_2\Bigg)s_{12}-{\rm Im}\big(\delta_3e^{-i(\theta_{12}+2\theta_{23})}\big)\Bigg] + \mathcal{O}(\epsilon^3) ~, \\
	{g}_{1113} &= \frac{3}{2}\epsilon\Bigg[\Bigg(\frac{2m_{h_1}^2}{v^2}-\delta_2\Bigg)c_{12}+{\rm Re}\big(\delta_3e^{-i(\theta_{12}+2\theta_{23})}\big)\Bigg] + \mathcal{O}(\epsilon^3) ~, \\
	{g}_{1223} &= \frac{\epsilon}{24v_s^2}\Big\{ 6\Big[-2(m_{h_2}^2+m_{h_3}^2)+(v^2+2v_s^2)\Big]c_{12}+36(m_{h_2}^2-m_{h_3}^2)c_{12+2(23)} \nonumber \\
	&\quad +12b_2(c_{12}+3c_{3(12)+4(23)}) -6{\rm Re}\big(3v^2\delta_3e^{-i(3\theta_{12}+4\theta_{23})}-3v_s^2\delta_3e^{i(3\theta_{12}+2\theta_{23})} \nonumber \\
	&\quad +v_s^2\delta_3e^{-i(\theta_{12}+2\theta_{23})}\big) -2\sqrt{2}v_s{\rm Re}\big( 9c_1e^{-i(3\theta_{12}+4\theta_{23})}-4c_2c_{12}-3c_2e^{i(3\theta_{12}+4\theta_{23})} \big) \nonumber \\
	&\quad -9v_s^2s_{12+23}\big[{\rm Im}\big( 2d_3c_{12+23}e^{i(\theta_{12}+2\theta_{23})}\big) +{\rm Im}\big(3d_3e^{-i(2\theta_{12}+3\theta_{23})}+2d_3e^{-i(2\theta_{12}+\theta_{23})} \nonumber \\
	&\quad +d_3e^{-i\theta_{23}}\big)\big] \Big\} + \mathcal{O}(\epsilon^2) ~, \\
	{g}_{1233} &= \frac{\epsilon}{24v_s^2}\Big\{ 6\Big[-2(m_{h_2}^2+m_{h_3}^2)+(v^2+2v_s^2)\Big]s_{12}-36(m_{h_2}^2-m_{h_3}^2)s_{12+2(23)} \nonumber \\
	&\quad +12b_2(s_{12}-3s_{3(12)+4(23)})-6{\rm Im}\big(3v^2\delta_3e^{-i(3\theta_{12}+4\theta_{23})}-3v_s^2\delta_3e^{i(3\theta_{12}+2\theta_{23})} \nonumber \\
	&\quad -v_s^2\delta_3e^{-i(\theta_{12}+2\theta_{23})}\big)-2\sqrt{2}v_s{\rm Im}\big( 9c_1e^{-i(3\theta_{12}+4\theta_{23})}-4c_2s_{12}+3c_2e^{i(3\theta_{12}+4\theta_{23})} \big) \nonumber \\
	&\quad -9v_s^2c_{12+23}\big[{\rm Re}\big( 2d_3s_{12+23}e^{i(\theta_{12}+2\theta_{23})} \big)+{\rm Im}\big(3d_3e^{-i(2\theta_{12}+3\theta_{23})}+2d_3e^{-i(2\theta_{12}+\theta_{23})} \nonumber \\
	&\quad -d_3e^{-i\theta_{23}}\big)\big] \Big\} + \mathcal{O}(\epsilon^2) ~, \\
	{g}_{1222} &= \frac{\epsilon}{4v_s^2}\Big\{ 3\Big[-2(m_{h_2}^2+m_{h_3}^2)+(v^2+2v_s^2)\delta_2\Big]s_{12}+6(m_{h_2}^2-m_{h_3}^2)s_{12+2(23)} \nonumber \\
	&\quad +6b_2(s_{12}+s_{3(12)+4(23)})+3{\rm Im}\big( v^2\delta_3e^{-i(3\theta_{12}+4\theta_{23})} - 2v_s^2\delta_3s_{12}e^{-i(2\theta_{12}+2\theta_{23})} \big) \nonumber \\
	&\quad +\sqrt{2}v_s{\rm Im}\big(3c_1e^{-i(3\theta_{12}+4\theta_{23})}+c_2e^{i(3\theta_{12}+4\theta_{23})}\big)+4\sqrt{2}v_s{\rm Re}(c_2s_{12}\big) \nonumber \\
	&\quad -3v_s^2s_{12+23}^2{\rm Im}\big( 2d_3e^{-i\theta_{12}}+3d_3e^{-i(\theta_{12}+2\theta_{23})}+d_3e^{i(\theta_{12}+2\theta_{23})} \big) \Big\} + \mathcal{O}(\epsilon^2) ~, \\
	{g}_{1333} &= \frac{\epsilon}{4v_s^2}\Big\{ 3\Big[-2(m_{h_2}^2+m_{h_3}^2)+(v^2+2v_s^2)\delta_2\Big]c_{12}-6(m_{h_2}^2-m_{h_3}^2)c_{12+2(23)} \nonumber \\
	&\quad +6b_2(c_{12}-c_{3(12)+4(23)})+3{\rm Re}\big( v^2\delta_3e^{-i(3\theta_{12}+4\theta_{23})} - 2v_s^2\delta_3c_{12}e^{-i(2\theta_{12}+2\theta_{23})} \big) \nonumber \\
	&\quad +\sqrt{2}v_s{\rm Re}\big(3c_1e^{-i(3\theta_{12}+4\theta_{23})} -c_2e^{i(3\theta_{12}+4\theta_{23})}\big)-4\sqrt{2}v_s{\rm Re}\big(c_2c_{12}\big) \nonumber \\
	&\quad +3v_s^2c_{12+23}^2{\rm Re}\big( 2d_3e^{-i\theta_{12}}+3d_3e^{-i(\theta_{12}+2\theta_{23})}-d_3e^{i(\theta_{12}+2\theta_{23})} \big) \Big\} + \mathcal{O}(\epsilon^2) ~.
\end{align}
where $s_{m(12)+n(23)}=\sin(m\theta_{12}+n\theta_{23})$, $c_{m(12)+n(23)}=\cos(m\theta_{12}+n\theta_{23})$.
\item
Couplings of Scalar Fields to SM Fermions and Gauge Bosons:
\begin{gather}	
h_1f\overline{f}: -\frac{m_f}{v}\left(1-\frac{\epsilon^2}{2}\right)
~,~
h_1WW: \frac{2m_W^2}{v}\left(1-\frac{\epsilon^2}{2}\right)
~,~
h_1ZZ: \frac{m_Z^2}{v}\left(1-\frac{\epsilon^2}{2}\right)
~, \\
h_2f\overline{f}: -\frac{m_f}{v}(-\epsilon s_{12})
~,~
h_2WW: \frac{2m_W^2}{v}(-\epsilon s_{12})
~,~
h_2ZZ: \frac{m_Z^2}{v}(-\epsilon s_{12})
~,  \\
h_3f\overline{f}: -\frac{m_f}{v}(-\epsilon c_{12}) 
~,~
h_3WW: \frac{2m_W^2}{v}(-\epsilon c_{12}) 
~,~
h_3ZZ: \frac{m_Z^2}{v}(-\epsilon c_{12}) 
~, 
\end{gather}
\end{itemize}

\section{Formulae for Electroweak Oblique Corrections}\label{sec:b}

The scalar contributions to $\Delta T$ and $\Delta S$ are given by
\begin{equation}
\label{eq:EWPO}
\begin{aligned}
	\Delta S^h &= \epsilon^2\Big[-\Delta S^h_{SM}(m_{h_1})+s_{12}^2\Delta S^h_{SM}(m_{h_2})+c_{12}^2\Delta S^h_{SM}(m_{h_3})\Big] ~, \\
		\Delta T^h &= \epsilon^2\Big[-\Delta T^h_{SM}(m_{h_1})+s_{12}^2\Delta T^h_{SM}(m_{h_2})+c_{12}^2\Delta T^h_{SM}(m_{h_3})\Big] ~,
\end{aligned}
\end{equation}
where
\begin{equation}
	\Delta S^h_{SM}(m_h) = -\frac{1}{\pi}\Bigg[\Big(1-\frac{x}{3}+\frac{x^2}{12}\Big)F(x) -\frac{3x-7}{72}\Bigg] ,~ x=m_h^2/m_Z^2 ~,
\end{equation}
\begin{equation}
	\Delta T^h_{SM}(m_h) = -\frac{3}{16s_W^2\pi}\Bigg[m_h^2\frac{\log(m_h^2/m_W^2)}{m_W^2-m_h^2}-\frac{m_Z^2}{m_W^2}m_h^2\frac{\log(m_h^2/m_Z^2)}{m_Z^2-m_h^2}\Bigg] ~,
\end{equation}
and
\begin{equation}
	F(x) = \begin{cases}
	\displaystyle
		1+\Bigg(\frac{x}{x-1}-\frac{1}{2}x\Bigg)\log x+x\sqrt{\frac{x-4}{x}}\log\Bigg(\sqrt{\frac{x-4}{4}}+\sqrt{\frac{x}{4}}\Bigg) ,~ x>4 \\
	\displaystyle
		1+\Bigg(\frac{x}{x-1}-\frac{1}{2}x\Bigg)\log x-x\sqrt{\frac{4-x}{x}}\tan^{-1}\sqrt{\frac{4-x}{x}} ,~ x<4
	\end{cases} ~.
\end{equation}

We have checked the consistency of these formulae with those in Ref.~\cite{Aguilar-Saavedra:2019ghg}. The only difference is in $\Delta S^h_{SM}$, the formula of which given in Ref.~\cite{Aguilar-Saavedra:2019ghg} is $\epsilon_3$ mentioned in Ref.~\cite{Barbieri:2004qk}, which further includes the derivative corrections.

%

\end{document}